\documentclass[useAMS, usenatbib]{mn2e}

\usepackage{graphicx}
\usepackage{lscape}
\usepackage{rotating}
\usepackage{scalefnt}
\usepackage{dblfloatfix}
\usepackage{multirow}
\usepackage{appendix}

\newif\ifAMStwofonts

\bibpunct{(}{)}{;}{a}{}{,}

\newcommand{\kms}{km~s$^{-1}$\,}

\newcommand {\hi} {H\,{\small I}\,}

\newcommand {\sL}{\rm\,L$_\odot$}

\newcommand {\vi} {{\it V--I\/}}
\newcommand {\mv} {{\it V\/}}
\newcommand {\mi} {{\it I\/}}

\title[Survey of the Phoenix transition type]{A wide-area view of the Phoenix 
dwarf galaxy from VLT/FORS imaging\thanks{Based on FORS observations collected at the ESO, proposal 083.B-0252.}}
\author[G.Battaglia et al.]{G.Battaglia$^{1, 2}$\thanks{E-mail:
gbattaglia@oabo.inaf.it}, M.Rejkuba$^{2}$, E.Tolstoy$^{3}$, M.J.Irwin$^{4}$, G.Beccari$^{2}$ \\
$^{1}$INAF - Osservatorio Astronomico di Bologna, via Ranzani 1, 40127 Bologna, Italy\\
$^{2}$European Organization for Astronomical Research in the Southern Hemisphere, K. Schwarzschild-Str. 2, 85748 Garching, Germany\\
$^{3}$Kapteyn Astronomical Institute, University of Groningen, P.O.Box 800, 9700 AV Groningen, the Netherlands \\
$^{4}$Institute of Astronomy, Madingley Road, Cambridge CB03 0HA, UK \\
}
\begin{document}

\date{Accepted 2012 May 10. Received 2012 April 18; in original form 2012 January 25}

\pagerange{\pageref{firstpage}--\pageref{lastpage}} \pubyear{2002}

\maketitle

\label{firstpage}

\begin{abstract}

We present results from a wide-area photometric survey of the Phoenix dwarf galaxy, one of the 
rare dwarf irregular/ dwarf spheroidal transition type 
galaxies (dTs) of the Local Group (LG). These objects  
offer the opportunity to study the existence of possible 
evolutionary links between the late- and early- type 
LG dwarf galaxies, since the properties of dTs suggest that they may 
be dwarf irregulars in the process of transforming into dwarf spheroidals. 

Using FORS at the VLT we have acquired VI photometry of Phoenix. 
The data reach a S/N$\sim$10 just below the horizontal branch of the system and 
consist of a mosaic of images that covers an area of 26\arcmin $\times$ 26\arcmin centered 
on the coordinates of the optical center of the galaxy. 

Examination of the colour-magnitude diagram and luminosity function
revealed the presence of a bump above the red clump, consistent with
being a red giant branch bump.

The deep photometry combined with the large area covered allows us to put 
on a secure ground the determination of the overall structural properties of the galaxy and 
to derive the spatial distribution of stars in different evolutionary phases and age ranges, 
from 0.1 Gyr to the oldest stars. 
The best-fitting 
profile to the overall stellar population is a Sersic profile of Sersic radius R$_S = $1.82\arcmin$\pm$0.06\arcmin 
and $m=$0.83$\pm$0.03.

We confirm that the spatial distribution of stars is found to become more and more centrally concentrated the 
younger the stellar population, as reported in previous studies.   
This is similar to the stellar population gradients found for close-by Milky Way dwarf spheroidal galaxies. 
We quantify such spatial variations by analyzing the 
surface number density profiles of stellar populations in different age ranges; the parameters of the best-fitting 
profiles are derived, and these can provide  
useful constraints to models 
exploring the evolution of dwarf galaxies in terms of their star formation. 

The disk-like distribution previously found in the central regions in Phoenix appears to 
be present mainly among stars younger than 1 Gyr, and absent for the stars $\ga$ 5 Gyr old, which on the other hand 
show a regular 
distribution also in the center of the galaxy. This argues against a disk-halo structure of the type 
found in large spirals such as the Milky Way.

\end{abstract}


\section{Introduction}
Dwarf galaxies are the most common type of galaxies in the nearby universe \citep[see e.g.][]{karachentsev2004}, 
and are expected to be among the first systems to form in the cosmological context. 
At larger redshifts galaxies such as those typically seen in the Local Group (LG) cannot be observed. 
Therefore detailed observations of current properties of LG galaxies offer the opportunity to gain insight  
into the earliest stages of galaxy formation in the most numerous type of galaxies in the Universe.

LG dwarf galaxies are classified in two main categories: the late-type objects, containing gas and currently forming stars,
 often have 
irregular appearance in the optical and therefore are called dwarf irregulars (dIrr), 
while the early-type dwarf spheroidals (dSph) - devoid of neutral gas -  
show a more regular morphology and no current star formation \citep[e.g.][]{mateo1998}. 
Dwarf galaxies with intermediate properties - such as no current star formation but containing
gas - are called transition types (dTs). 

An open question is whether the
various types of dwarf galaxies in the LG are intrinsically different objects, or whether 
they descend from the same progenitors and have evolved through different 
paths because of environmental and/or internal processes.

The continuum of properties shown by the various types of LG dwarf galaxies \citep{tolstoy2009}, which makes their classification 
sometimes uncertain, seems to argue in favour of an evolutionary link. For example, PegDIG is the most luminous of the few LG 
transition type dwarfs (Phoenix, LGS3, DDO210, Leo~T, VV~124) and, because of this, it is sometimes considered as a dIrr. 
For Fornax, which has formed stars until very recently \citep[50-100 Myr ago][]{coleman2008} but is now devoid of gas, its classification as a dSph 
seems to be merely the result of the particular moment in time in which we are observing this galaxy: 
if we had been able to observe Fornax a few hundreds Myr ago, while still forming stars, 
we would have classified it as a dT or even as a dIrr. 

Another hint to a possible evolutionary link is the 
existence of a ``morphology-density'' relation: dIrrs are found at relatively large distances from the large LG spirals 
(i.e. $>$ 300 kpc from the Milky Way and M31), while the great majority of early type dwarf galaxies are satellites of the Milky Way or M31, being located at 
$<$ 300 kpc from them. This already suggests that the interaction with the large
spirals (hereafter referred to as ``environment'') must have played a role in determining the different evolutionary
paths of late and early type dwarfs. Models based on N-body simulations do show that such interactions are a
viable explanation to the observed ``morphology-density'' relation \citep[e.g.][]{mayer2006, kazantzidis2011}. 
However, the presence of dSphs (e.g. Cetus and Tucana) and dTs (e.g. Aquarius) at distances such that any
strong interaction with either the MW or M31 can be excluded indicates that the environment cannot be the
only factor at play but that also internal factors (e.g. the depth of the potential well of the single objects) might
play a role.

The lack of knowledge about the detailed properties of dIrrs and dTs hinders the 
exploration of a possible common origin of LG dwarfs, and what might have caused a different 
evolutionary path.

At a distance of 415 kpc \citep[][hereafter H09]{hidalgo2009}, Phoenix is the closest transition type dwarf, but even 
basic properties such as its extent are rather uncertain. For example,  \citet[][hereafter VDK91]{vdk1991} 
derived an extent of 8.7\arcmin, while \citet[][hereafter MD99]{martinez1999} 
found a larger value, with a nominal tidal radius of $r_t = 15.8_{-2.8}^{+4.3}$ arcmin. 
Also its heliocentric 
systemic velocity is under debate: \citet{irwin2002} derived a velocity of $-13\pm 9$ \kms, while \citet{gallart2001} 
measured  
a velocity of $-52 \pm 6$ \kms. Both studies though 
consider likely an association between Phoenix and the cloud of \hi\, gas found around the galaxy at heliocentric 
velocity between $-30$ and $-14$ \kms \citep{young2007}.  

The star formation history (SFH) of Phoenix has instead been extensively studied  
\citep[see][]{vdk1991, held1999, martinez1999, holtzman2000, gallart2004, menzies2008, hidalgo2009}. 
The co-existence of classical Cepheids and anomalous Cepheids and RR Lyrae variables in this system argues for an extended 
and complex star formation history \citep{gallart2004}. 
Phoenix has a similar luminosity to the Sculptor dSph ($L_V = 0.9 \times 10^6 L_\odot$ and $L_V = 2.15 \times 10^6 L_\odot$, 
respectively; Mateo 1998), and like Sculptor it formed most of its stars more than 10 Gyr ago (H09). 
However, unlike Sculptor or the majority of other dSphs it does show recent star formation (MD99), 
possibly with stars as young as 100 Myr, and there is HI gas associated with the system \citep{gallart2001}. This means 
that, unlike the similarly luminous Sculptor, this system was able to retain its gas for the whole of its lifetime, be it 
because of a larger potential well and/or a less harsh environment, with less 
stripping and a larger possibility of gas infall with respect to a satellite of the MW.

The most recent study of Phoenix SFH (H09) 
is based on two HST pointings that, even if covering a small fraction of the galaxy, 
reach out to about 4\arcmin from the center, giving a view of how the SFH changed across the 
object. The central regions contain stars over a range of ages (from 100 Myr to the oldest stars), 
while already at 450pc from the center 95\% of the stars are more than 8 Gyr old. 
These properties appear to be 
consistent with the star formation region slowly shrinking with time (H09). 

In this paper we investigate the global properties of Phoenix based on wide-area 
imaging data collected at the ESO VLT with FORS. The observations cover a large region, 26\arcmin$\times$26\arcmin, and 
reach below the horizontal branch of the galaxy, 
containing stars in evolutionary stages representative of the whole age range displayed by Phoenix. This 
allows us not only to explore the overall structure of the galaxy but also 
to directly derive the spatial distribution of stars in different age ranges. The paper is organised as 
follows: in Sect.~\ref{sec:obs} we present the observations and the data reduction procedure adopted; 
in Sect.~\ref{sec:str} we analyze the overall structure of the galaxy; in Sect.~\ref{sec:tip} 
we use the tip of the RGB to derive an estimate of the distance modulus to Phoenix; 
Sect.~\ref{sec:variations} deals with the stellar population mix and its spatial variations; we conclude 
with a discussion in Sect.~\ref{sec:disc} and a summary of the results in Sect.~\ref{sec:concl}.

In a subsequent paper we will complement this photometric study with CaT spectroscopy 
to investigate the metallicity and stellar population gradients in this galaxy.

\begin{table*}
\begin{center}
\begin{tabular}{lccc}
\hline
\hline
Parameter & value & reference & notes\\
\hline
($\alpha_{\rm J2000}$,$\delta_{\rm J2000}$) &  01$^h$ 51$^m$ 06$^s$, $-$44\degr 26\arcmin 42\arcsec & 1 & \\
P.A. &  8\degr $\pm$ 4\degr & 2 & beyond 4.5\arcmin\\
$e$  &  0.3$\pm$0.03 & 2 & beyond 4.5\arcmin \\
R$_{\rm core}$ & 1.79\arcmin$\pm$0.04\arcmin & 2 & \\
R$_{\rm tidal}$ & 10.56\arcmin$\pm$0.15\arcmin & 2 & \\
R$_{\rm 1/2}$ &  2.30\arcmin$\pm$0.07\arcmin & 2 & \\
(m-M)$_0$  & 23.06$\pm$0.12   & 2 & \\
Distance  & 409$\pm$23 kpc & 2 & \\
L$_V$ &  $0.9 \times 10^6$ \sL & 1 & referred to a distance of 445 kpc \\
V$_{\rm HB}$ & 23.9 & 3 & \\
E(B-V)       & 0.016 & 4 & \\
A$_V$ & 0.050 mag &  &  \\
A$_I$ & 0.024 mag &  &  \\
1 arcmin/kpc    &  0.119  & & assuming a distance of 409 kpc\\ 
\hline
\end{tabular}
\caption{The various rows are from top to bottom: coordinates of the optical centre; 
position angle P.A., defined as the angle between the north direction and the major axis of the galaxy measured 
counter-clockwise; ellipticity, defined as $e=1 - b/a$; King core and tidal radius; half-light radius, 
defined as $R_{1/2} \sim 1.68 \times \mathrm{R}_{\rm e}$, where R$_{\rm e} = $ 1.37$\pm$0.01\arcmin\, is the exponential radius; 
 distance modulus and heliocentric distance; luminosity in V-band; V magnitude of the 
horizontal branch; reddening; extinction in \mv and \mi band. The last row quotes the conversion factor 
from arcmin to kpc at the distance of Phoenix. The P.A. and ellipticity refer to the outer population, aligned north-south. 
References: 1 $=$ \citet{mateo1998}; 2 $=$ this work; 3 $=$ \citet{holtzman2000}; 
4 $=$ \citet{schlegel1998}, http://irsa.ipac.caltech.edu/applications/DUST/, average value over 2 degrees. The 
$A_V$ value assumes $A_V/$E(B-V) $=$3.1; A$_I$ was derived using the ratio A$_I$/A$_V = 0.479$ from \citet{cardelli1989}; note 
that previous works such as H09 assume larger values of the reddening, 0.02. 
} \label{tab:par}
\end{center}
\end{table*}

\begin{figure}
\begin{center}
\includegraphics[width=90mm]{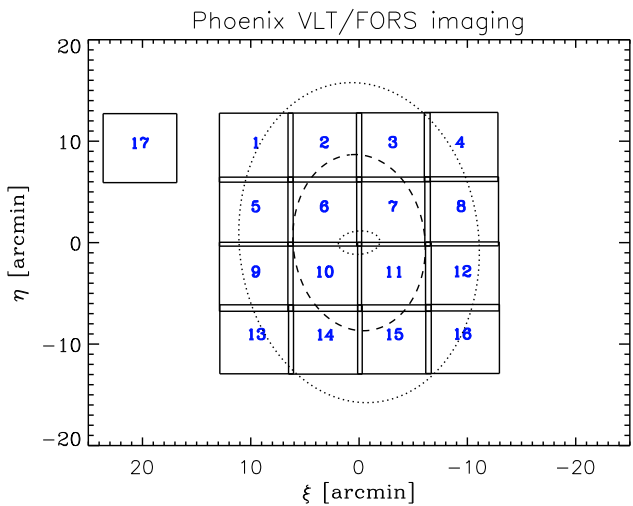}
\caption{Schematic representation of the coverage of our VLT/FORS photometric data for the Phoenix dwarf galaxy. The black squares 
indicate the individual pointings, showing the regions of overlap. The inner dotted ellipse (with parameters 
$e=0.4$, P.A.$=$ 95\degr and semi-major axis $=$ 115\arcsec) denotes the region 
where the elliptical isopleths show a sharp rotation in their major axes in the analysis of \citet{martinez1999}, 
and where most of the young stars are found. The outer ellipses indicate two measurements 
for the extent of Phoenix: the dotted one (with parameters $e=0.3$, P.A.$=$ 5\degr and semi-major axis $=$ 15.8\arcmin) shows the nominal tidal 
radius from the study of \citet{martinez1999}, and 
the dashed one (with parameters $e=0.3$, P.A.$=$ 5\degr and semi-major axis $=$ 8.7\arcmin) where the exponential fit to the surface brightness profile  
approaches zero from \citet{vdk1991}.}
\label{fig:cov_img}
\end{center}
\end{figure}

\begin{table*}
\begin{center}
\begin{tabular}{lccrcccc}
\hline
\hline
 Field name  &  RA(deg) & DEC(deg) & Date and UT of start of observation & Filter & exptime [sec] & airmass & seeing [arcsec] \\
\hline
phoenix-img01 &27.996042 & -44.28881 & 2009-07-27  9:33:14.097&         V   &  3$\times$120&  1.073   &  0.60    \\ 
              &  &                   &             9:41:03.332      &  I     & 5$\times$ 90& 1.069   &   0.60        \\
phoenix-img02 & 27.848583 & -44.28881& 2009-07-27 10:00:29.563 &       V    &  3$\times$120& 1.062    &  0.53        \\ 
              &  &                   &            10:08:24.488   &    I      & 5$\times$ 90&  1.062   &  0.50         \\
phoenix-img03 & 27.701458 & -44.28839& 2009-07-28  8:25:56.322 &       V    &  3$\times$120& 1.138    &  0.70        \\ 
              &  &                   &             8:33:59.459   &    I      & 5$\times$ 90&  1.126   &  0.63         \\
phoenix-img04 & 27.555667 & -44.28756& 2009-07-28  8:46:01.520 &       V    &  3$\times$120& 1.109    &  0.67        \\ 
              &  &                   &             8:53:49.835   &    I      & 5$\times$ 90&  1.100     &  0.63            \\
phoenix-img05 & 27.99625  & -44.39389& 2009-07-28  9:05:35.354  &      V    &  3$\times$120& 1.091   &   0.70        \\ 
              &  &                   &             9:13:23.819   &    I      & 5$\times$ 90&  1.084    &  0.61           \\
phoenix-img06 & 27.848833 & -44.3939 & 2009-07-24  8:28:10.733  &      V    &  3$\times$120& 1.163   &    0.73       \\ 
              &  &                   &             8:36:03.667   &    I      & 5$\times$ 90&  1.149    &  0.58        \\
phoenix-img07 & 27.700833 & -44.39395& 2009-07-24  8:03:08.061  &      V    &  3$\times$120& 1.216    &   0.63       \\ 
              &  &                   &             8:15:40.713   &    I      & 5$\times$ 90&  1.187   &   0.56        \\
phoenix-img08 & 27.554083 & -44.39334& 2009-07-28  9:30:56.350 &       V    &  3$\times$120& 1.071    &   0.65       \\ 
              &  &                   &             9:38:44.785   &    I      & 5$\times$ 90&  1.068   &   0.59        \\
phoenix-img09 & 27.996792 & -44.50086& 2009-07-28  9:51:22.636 &       V    &  3$\times$120& 1.065   &    0.66       \\ 
              &  &                   &             9:59:12.151   &    I      & 5$\times$ 90&  1.064   &   0.60        \\
phoenix-img10 & 27.849125 & -44.50111& 2009-07-24  8:48:37.538 &       V    &  3$\times$120& 1.130    &    0.66       \\ 
              &  &                   &             8:56:41.984   &    I      & 5$\times$ 90&  1.119   &   0.60        \\
phoenix-img11 & 27.69875 &  -44.50116& 2009-06-21  9:35:54.111 &       V    &  3$\times$120& 1.324    &   0.63       \\ 
              &  &                   &             9:43:45.942   &    I      & 5$\times$ 90&  1.298   &   0.63        \\
phoenix-img12 & 27.552167 & -44.50031& 2009-07-29  9:16:05.726  &      V    &  3$\times$120& 1.079   &    0.85       \\ 
              &  &                   &             9:23:58.611   &    I      & 5$\times$ 90&  1.074    &  0.84        \\
phoenix-img13 & 27.996333 & -44.60359& 2009-07-29  9:35:54.979 &       V    &  3$\times$120& 1.070    &    0.73       \\ 
              &  &                   &             9:43:42.634   &    I      & 5$\times$ 90&  1.067    &  0.69        \\
phoenix-img14 & 27.847792 & -44.6043 & 2009-07-29  9:56:11.015 &       V    &  3$\times$120& 1.064   &    0.83       \\ 
              &  &                   &            10:04:22.331   &    I      & 5$\times$ 90&  1.064   &   0.80        \\
phoenix-img15 & 27.698542 & -44.60388& 2009-08-20  6:24:45.464 &       V    &  3$\times$120& 1.199   &    0.83       \\ 
              &  &                   &             6:32:43.001   &    I      & 5$\times$ 90&  1.182   &   0.75        \\
phoenix-img16 & 27.551292 & -44.60303& 2009-08-20  6:45:15.295 &       V    &  3$\times$120& 1.157    &   0.86       \\ 
              &  &                   &             6:53:06.891   &    I      & 5$\times$ 90&  1.143    &  0.80        \\
phoenix-img17 & 28.246042 & -44.28881& 2009-08-20  7:05:48.935 &       V    &  3$\times$120& 1.126    &   0.84       \\ 
              &  &                   &             7:13:51.822   &    I      & 5$\times$ 90&  1.115    &  0.64          \\
\hline
\end{tabular}
\caption{Table of FORS imaging observation for the Phoenix dT. The seeing is the average stellar 
FWHM from the final, combined, image. 
} \label{tab:journal_phx_img}
\end{center}
\end{table*}

\begin{table*}
\begin{center}
\begin{tabular}{lrcccc}
\hline
\hline
 Field name  & Date and UT of observation & Filter & airmass & exptime [sec] & seeing [arcsec] \\
\hline
       E5 &  2009-06-20  22:42:09.728 &  V &   1.071 &   3 & 0.53 \\
          &              22:43:39.426 &  I &   1.071 &   1 & 0.45\\
       E7 &  2009-07-24  02:02:19.507 &  V &   1.067 &   3 & 0.80 \\
          &              02:03:50.096 &  I &   1.067 &   1 & 1.00\\
      L92 &  2009-07-27  10:27:40.238 &  V &   1.163 &   3 & 0.50 \\
          &              10:29:05.406 &  I &   1.165 &   1 & 0.53\\
    MarkA &  2009-07-27  05:58:03.331 &  V &   1.055 &   3 & 0.45\\
          &              05:59:28.549 &  I &   1.056 &   1 & 0.43\\
 PG0231\_1 &  2009-07-24  10:23:00.265 &  V &   1.176 &   3 & 0.65\\
           &              10:24:28.883 &  I &   1.175 &   1 & 0.75\\
 PG0231\_2 &  2009-07-28  10:13:18.350 &  V &   1.171 &   3 & 0.60 \\
           &              10:14:43.538 &  I &   1.169 &   1 & 0.65\\
   PG1525  &  2009-06-20  22:50:50.644 &  V &   1.597 &   3  & 0.55\\
           &              22:52:19.402 &  I &   1.585 &   1  & 0.50 \\
   PG1657  &  2009-07-27  00:32:21.166 &  V &   1.213 &   3  & 0.53\\
          &               00:33:46.404 &  I &   1.212 &   1  & 0.43\\
 PG2213\_1 &  2009-07-27  07:10:52.208 &  V &   1.108 &   3 & 0.50\\
           &              07:12:17.476 &  I &   1.109 &   1 & 0.48\\
 PG2213\_2 &  2009-07-28  10:22:26.352 &  V &   2.015 &   3 & 0.73\\
           &              10:23:51.520 &  I &   2.035 &   1 & 0.65\\
 PG2213\_3 &  2009-07-29  06:47:34.819 &  V &   1.100 &   3 & 0.50\\
           &              06:49:00.078 &  I &   1.101 &   1 & 0.45\\
 PG2213\_4 &  2009-07-29  09:11:13.858 &  V &   1.437 &   3 & 0.78\\
           &              09:12:39.046 &  I &   1.445 &   1 & 1.00\\
 PG2213\_5 &  2009-07-29  10:29:37.794 &  V &   2.181 &   3 & 1.50\\
           &              10:31:03.072 &  I &   2.204 &   1 & 0.88\\
 PG2213\_6 &  2009-08-20  07:28:30.579 &  V &   1.359 &   3 & 0.78\\
           &              07:29:09.972 &  I &   1.362 &   1 & 0.70\\
  T\_Phe\_1 &  2009-07-28  09:27:24.530 &  V &   1.088 &   3 & 0.60 \\
            &             09:28:49.538 &  I &   1.089 &   1 & 0.43\\
  T\_Phe\_2 &  2009-07-29  10:23:12.828 &  V &   1.149 &   3 & 0.75\\
            &             10:24:38.096 &  I &   1.151 &   1 & 1.38 \\
  T\_Phe\_3 &  2009-08-20  07:33:51.860 &  V &   1.079 &   3 & 0.68\\
            &              07:34:30.794 &  I &   1.079 &   1 & 0.60 \\
\hline
\end{tabular}
\caption{Table of FORS imaging observation for standard fiels. The seeing is the average stellar 
FWHM from the final, combined, image. 
} \label{tab:journal_std_img}
\end{center}
\end{table*}

\section{Observations and Data Reduction}\label{sec:obs}
The observations were carried out between June 20 and August 20, 2009,  
in service mode with the ESO VLT instrument FORS2 at UT1 (Antu). FORS2 is a focal reducer multi mode instrument that can be used for optical imaging, polarimetry, longslit and multi-object spectroscopy \citep{appenzeller1992}. It has field of view of 6\farcm 8 $\times$ 6\farcm 8 with pixel size 0\farcs 25. The detectors are two $2k \times 4k$ MIT CCDs 
(hereafter referred to as C1 and C2). We used FORS2 in imaging mode to make a mosaic of 4$\times$4 fields covering 
an area of approximately 
26 \arcmin $\times$ 26 \arcmin 
centered on the coordinates of the optical centre of Phoenix 
(see Table~\ref{tab:par} for the galaxy parameters). One additional field was observed at a displaced 
location from the galaxy as a check on the determination of the foreground/background density 
(see Fig.~\ref{fig:cov_img}). For 
each pointing we took 3 dithered exposures of 120s each in $v_{\rm HIGH}$\footnote{This filter has a 
central wavelength of 555nm and a FWHM of 123.2; for the transmission curve see 
http://www.eso.org/sci/facilities/paranal/instruments/fors/doc/VLT-MAN-ESO-13100-1543\_v87.pdf} and 5 dithered exposures of 90s each in $I_{\rm Bess}$ 
filter (in the subsequent text for brevity we denote filters as V instead of v$_{\rm HIGH}$ and I instead of 
I$_{\rm Bess}$).  The offset sizes between the exposures were such that we covered the gaps between the two detectors.  Table~\ref{tab:journal_phx_img} lists the observation log of all scientific fields.

As a part of the standard calibration plan for FORS2, observations of standard star fields in these filters
were also carried out by ESO service mode observers. 
The log of all standard star observations is in 
Table~\ref{tab:journal_std_img}.

The basic data reduction steps, consisting of bias subtraction and division by the normalized twilight flat field exposure, 
were performed with the ESO FORS imaging pipeline.

For each pointing such pre-reduced images were first aligned to the coordinates 
of one of the individual exposures, which was acting as the reference image. 
The regions of the CCD that had no data were masked out  - in particular the so-called slave CCD (bottom CCD) has a large area that gets no light - and the aligned images were then median combined using IRAF task {\it imcombine}. 
The image quality was measured on combined images
and the FWHM values are given in Table~\ref{tab:journal_phx_img}.  We checked that the FWHM 
measured on the combined images was similar to those in the individual exposures. 

Since the internal regions of Phoenix are relatively crowded, 
we decided to perform PSF fitting photometry on all of our pointings to get uniform photometry over the whole area. 
This was done using the standalone version of the DAOPHOT~{\small II} package \citep{stetson1987}.

The routines FIND + PHOTOMETRY + PICK + PSF + ALLSTAR were run separately for each photometric band and for the two FORS CCD chips, 
accounting for the slightly different read-out noise of the CCD chips. 

Default parameters were used for the FIND routine to deal with the rejection of bad pixels and elongated objects along the direction of rows and columns. First, objects were located by imposing a 3$\sigma$ and 4$\sigma$ threshold above the 
background on the individual images for the V and I band, respectively. This threshold value was chosen after inspecting the number of detected objects as a function of threshold parameter, and selecting the value corresponding to the ``knee'' where the number of detected objects is changing rapidly with small changes in $\sigma$ for threshold. 

The choice of the PSF stars was done interactively for each field, separately for each filter and each CCD. 
First the routine PICK was run to automatically select a sample of unsaturated, relatively isolated stars per frame; 
a PSF was then created, and an image containing the residuals using ALLSTAR. 
All the PSF stars were then visually examined both on the original image and on the image with subtracted stars. 
Galaxies, blends or stars with close companions were manually removed from the sample, and the PSF re-determined using this subsample of stars. 
The procedure was repeated until the only remaining stars to be used for the PSF determination were non-saturated and relatively isolated stars. 
Typically,  
between 15 and 30 such stars per frame were 
used to determine the PSF. We used 
the option for a spatially non variable PSF, after testing that no significant differences 
were introduced by allowing for a spatially variable PSF. The analytical first approximation used 
as model for the PSF was a Moffat function.

The above procedure, except the creation of the PSFs, was then re-run 
on the residual images, in order to extract stars that  
were not previously found due to the wings of brighter stars. We checked that 
this procedure was not fitting noise by overplotting the position of these fainter  
stars on top of the image of residuals.

Finally, DAOMATCH and DAOMASTER were used to combine V and I ALLSTAR PSF fitting photometry of each pointing - 
separately for the 
2 CCD chips. Only objects detected in both bands and with magnitude  
errors $<$ 0.5 mag were retained.

Aperture corrections were computed for each field, chip and filter separately using aperture photometry 
and a curve-of-growth method on a set of well exposed and isolated PSF stars.

\subsection{Astrometry and Photometric calibration} \label{sec:cal}
The J2000 celestial coordinates of the detected objects were derived 
with CataXcorr\footnote{This code was developed by Montegriffo at INAF- 
Osservatorio Astronomico di Bologna, with the aim of cross-correlating catalogues and finding 
astrometric solution. The code is routinely used in publications from staff of the INAF- 
Osservatorio Astronomico di Bologna \citep[e.g.][]{bellazzini2011}.}. The 
astrometric solution was found by fitting typically a polynomial of third degree between 
at least 46 stars per ponting and the reference catalogue (USNOA2 and GSC2). The average 
r.m.s. of the solution was 0.2\arcsec in both RA and DEC.

Given that the standard fields used are not crowded, we performed aperture photometry 
rather than PSF photometry.  We located only luminous standard stars, 
by adopting a threshold of $15 \sigma$ above the background on the individual images
\footnote{We checked that reducing this threshold would not significantly increase the 
number of detected stars with catalogued magnitude per frame nor the quality of our calibration.}. 
The aperture at which to calculate the instrumental magnitude was
 derived by examining the run of the magnitude as a function of 
aperture size for those bright stars that are 
neither saturated, nor close to the border of the image, and for which 
the curve of growth reaches a constant value. 

In general an aperture of 25 pixels radius was found to 
include all the stars' flux for most of the fields of standard stars. However, 
this aperture radius was too large for a relatively crowded field such as E7; 
furthermore only a handful of stars passed the above criteria for 
fields E5 and PG0231: in these three cases we therefore used a smaller aperture radius 
and derived the aperture correction choosing appropriate isolated stars manually.

Applying the above described procedure, we extracted  
13 standard stars for the night of June 21 (9 chip1 + 4 chip2); 91 (61 + 30), 58 (36+22), 
19 (11+8), 28 (19+9) for July 24,27,28,29 respectively; 22 (12+10) for August 20. 

We note that the data available for the standard stars are not sufficient to fit simultaneously 
the zero point, the extinction coefficient and the color term for each band, night and chip separately. 
We therefore follow an iterative approach \citep{jerjen2001}, which we apply to the two CCD chips 
separately. 

For each band, we use the equation   
\begin{equation} \label{eq:calibr}
m_{\rm instr} = M + Z_m + k_m \times X_m + c_m \times (V-I)
\end{equation}
where $m_{\rm instr}$ and $M$ are the instrumental and calibrated 
magnitude, $X_m$ the airmass of the pointing, Z$_m$ the zero point, 
 k$_m$ the extinction coefficient and $c_m$ the color term. 
The calibrated photometry for the standard stars was extracted from the online catalogues 
of \citet{stetson2000}\footnote{http://www3.cadc-ccda.hia-iha.nrc-cnrc.gc.ca/community/STETSON/standards/}. 

On a first pass, we derive the zero point and extinction coefficient for each night 
separately by fitting the instrumental magnitudes and keeping 
the colour term fixed to zero. The derived coefficients are then 
applied to the calibrated magnitudes of the respective nights, 
obtaining $m_{\rm step1} = M + Z_{\rm m, fit} + k_{\rm m, fit} \times X_m$; 
the data of all the nights are then joined together to increase the statistics, and a color 
term is fitted to $m_{\rm instr} - m_{\rm step1} = c_{m, fit} \times (V-I) + const$. 
The fitting procedure of the zero point and extinction term night by night is then repeated, this time by keeping 
the color term fixed to $c_{m, fit}$. Any additional dependency on the color has been added to the color term.

\begin{table}
\begin{center}
\begin{tabular}{|l|cccccc|}
\hline
\hline
Night \& CCD chip & ZP V-band & ZP  I-band \\
\hline
21 Jun C1 & -27.934$\pm$ 0.025 &   -27.422$\pm$ 0.043  \\
21 Jun C2 & -27.997$\pm$ 0.009 &   -27.400$\pm$ 0.044  \\
24 Jul C1 & -28.035$\pm$ 0.007 &   -27.418$\pm$ 0.007  \\
24 Jul C2 & -28.022$\pm$ 0.007 &   -27.410$\pm$ 0.011  \\
27 Jul C1 & -27.956$\pm$ 0.038 &   -27.485$\pm$ 0.042  \\
27 Jul C2 & -28.027$\pm$ 0.010 &   -27.394$\pm$ 0.008  \\
28 Jul C1 & -28.012$\pm$ 0.030 &   -27.415$\pm$ 0.027  \\
28 Jul C2 & -28.049$\pm$ 0.017 &   -27.373$\pm$ 0.023  \\
29 Jul C1 & -28.080$\pm$ 0.025 &   -27.407$\pm$ 0.019  \\
29 Jul C2 & -28.032$\pm$ 0.026 &   -27.393$\pm$ 0.033  \\
20 Aug C1 & -28.005$\pm$ 0.034 &   -27.378$\pm$ 0.025  \\
20 Aug C2 & -28.049$\pm$ 0.049 &   -27.439$\pm$ 0.051  \\
\hline
\end{tabular}
\caption{Zero-points (and their errors) of the photometric calibration transformations (Eq.~\ref{eq:calibr}) 
derived for each night of observations, 
separated per band and CCD chip. 
} \label{tab:calibr}
\end{center}
\end{table}

\begin{figure*}
\includegraphics[width=\linewidth]{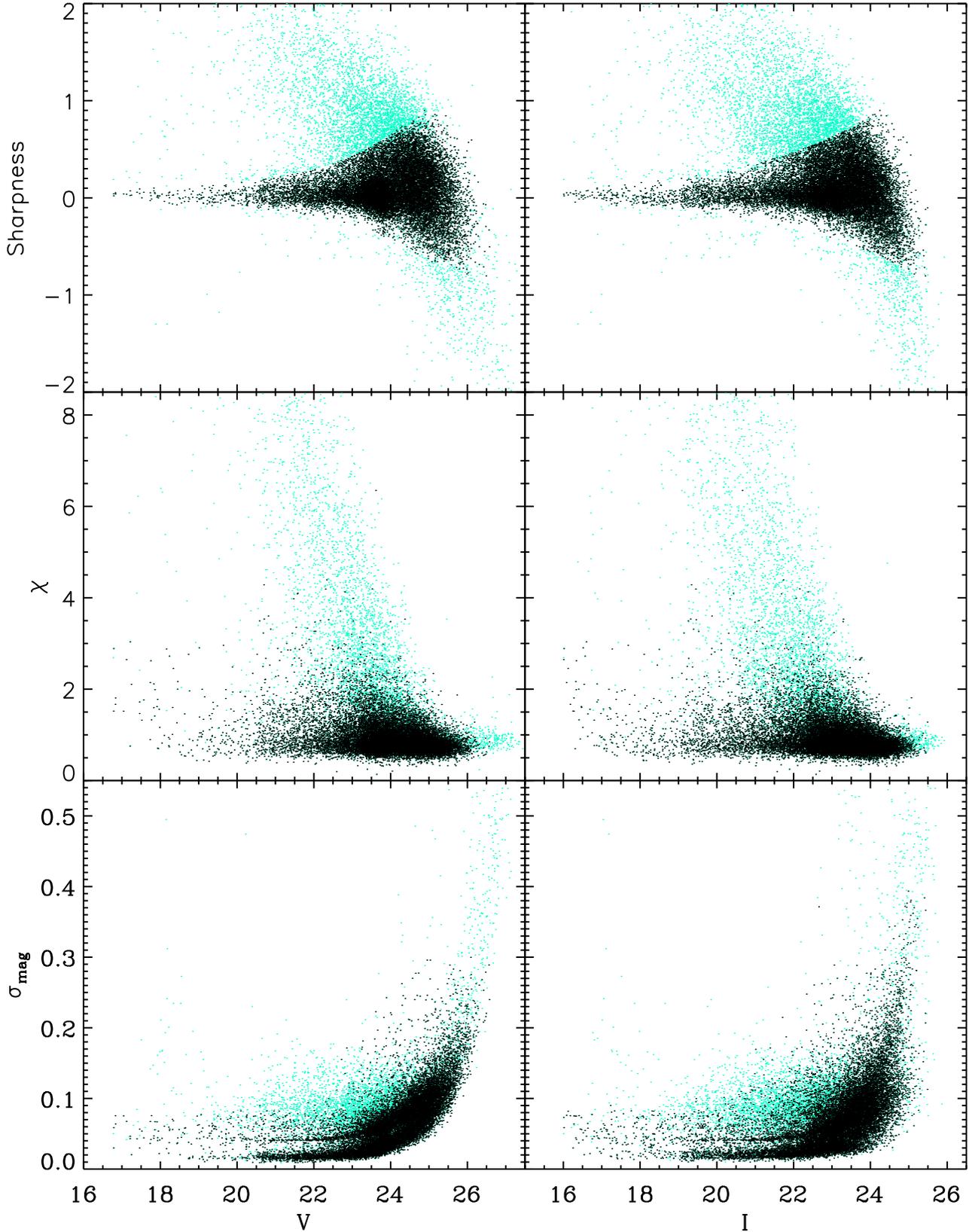}
\caption{Photometric parameters for the detected sources (cyan points) 
as a function of the calibrated V- (left) and I-(right) magnitudes. 
The top and central panels show the sharpness and $\chi$ parameters, respectively, from DAOPHOT. The bottom panels 
show the photometric errors. Beside the magnitude errors output from DAOPHOT, these errors factor in 
the uncertainties from PSF correction, photometric calibration and the shifts applied to the 
individual magnitues so as to bring all the photometry to the system of pointing 6. The black points  
show only those sources which we consider as ``stars'' on the basis of their sharpness parameter, and 
that have $\sigma_V < 0.3$ mag and $\sigma_I < 0.4$ mag. \label{fig:sharp}}
\end{figure*}

\begin{table*}
\begin{center}
\begin{tabular}{lcccccccc}
\hline
\hline
Name & RA [deg] & DEC [deg] & V & $\sigma_V$ & I & $\sigma_I$ & $\chi$ & sharpness \\ 
\hline
phx00001  &   27.9264897 & -44.2962170  & 25.0495 &  0.1396  &  23.8995  &   0.1396  &   1.2195  &   0.6250 \\
phx00002  &   28.0484214 & -44.2960092  & 25.4608 &  0.1679  &  23.8604  &   0.1679  &   1.1445  &   0.6710 \\
phx00003  &   28.0380659 & -44.2959729  & 25.3526 &  0.0983  &  22.8875  &   0.0983  &   0.7605  &   0.2435 \\
phx00004  &   28.0376092 & -44.2955426  & 24.8144 &  0.0926  &  24.5143  &   0.0926  &   0.6230  &   0.2570 \\
phx00005  &   28.0383987 & -44.2954079  & 25.3673 &  0.1125  &  23.0559  &   0.1125  &   0.7450  &   0.1105 \\
phx00006  &   27.9966041 & -44.2952897  & 24.2082 &  0.0921  &  23.2352  &   0.0921  &   1.4590  &   0.5680 \\
phx00007  &   27.9471308 & -44.2953097  & 24.5857 &  0.0768  &  24.0239  &   0.0768  &   0.7310  &   0.1025 \\
phx00008  &   28.0220882 & -44.2952300  & 25.6432 &  0.1224  &  23.3985  &   0.1224  &   0.9160  &   0.4125 \\
phx00009  &   27.9638667 & -44.2951593  & 24.3899 &  0.0847  &  23.6765  &   0.0847  &   0.9075  &   0.3655 \\
phx00010  &   28.0697991 & -44.2950072  & 24.5369 &  0.0996  &  23.6116  &   0.0996  &   1.2930  &   0.5940 \\
\hline
\end{tabular}
\caption{Catalogue of bona-fide stars from FORS photometry of the Phoenix dT, including pointing 17. The columns list the name 
of the star (1), RA (2) and DEC (3) in degrees, V magnitude (4), error in V magnitude (5), 
I magnitude (6), error in I magnitude (7), $\chi$ (8) and sharpness (9) parameters (8). {\it This is a sample of the table; the complete version will be published online.}
} \label{tab:catal}
\end{center}
\end{table*}

The average color term in V-band is 
 -0.063$\pm$0.012 for C1 and -0.053$\pm$0.006 for C2, 
and is consistent with zero for the I-band, 
0.010$\pm$0.006 for C1 and 0.001$\pm$0.01 for C2. This compares well with the color terms available 
on the ESO quality control webpages 
for P83\footnote{http://www.eso.org/observing/dfo/quality/FORS2/qc/photcoeff/ 
photcoeffs\_fors2.html; note that, 
as stated on this webpage, the errors 
on the tabulated quantities are underestimated.}, when available for the color used here.   
The average extinction term in V-band is 0.141$\pm$0.015 for C1 and 0.134$\pm$0.008 for C2, while in 
I-band it is 0.055$\pm$0.011 for C1 and 0.039$\pm$0.010 for C2; the determinations for the two chips are 
consistent with each others within 1$\sigma$. 
The nightly zero points are listed in Table~\ref{tab:calibr}. 
The calibrated magnitudes for the science targets were finally derived by iteratively applying 
the calibration transformations to the instrumental magnitudes, to which 
aperture corrections had been previously applied. The errors on the derived aperture corrections and  
coefficients of the calibration transformation are included in the error on the calibrated magnitudes.

We used the region of overlap between adjacent fields to place 
our photometry on a common photometric scale: for this we applied 
the magnitude shifts derived by comparing the magnitudes of objects 
in the overlap regions to the various pointings and tied 
all of them to the system of pointing 06. The applied shifts were typically $< \pm 0.1$mag; 
in part the magnitude shifts applied are the consequence of the errors in the calibration of the standard 
fields, but mostly they arise from the imperfect flat-fielding due to sky concentration in FORS 
\citep{freudling2007}\footnote{Although the report from Freudling et al. described in detail the photometric accuracy of FORS1, it is expected that FORS2 flat-fields present similar effects as the two cameras have the same optics and differ only in the detector type}. 
The errors in the determinations of the magnitude shifts were propagated onto the 
magnitude errors of the individual objects. Obviously pointing 17, which has 
no overlap with the main mosaic region, could not be placed on the same 
photometric system. Since however we use this pointing only for checks on  
the determination of the foreground/background density, this will not 
affect our analysis. 

\begin{figure*}
\includegraphics[width=0.7\linewidth]{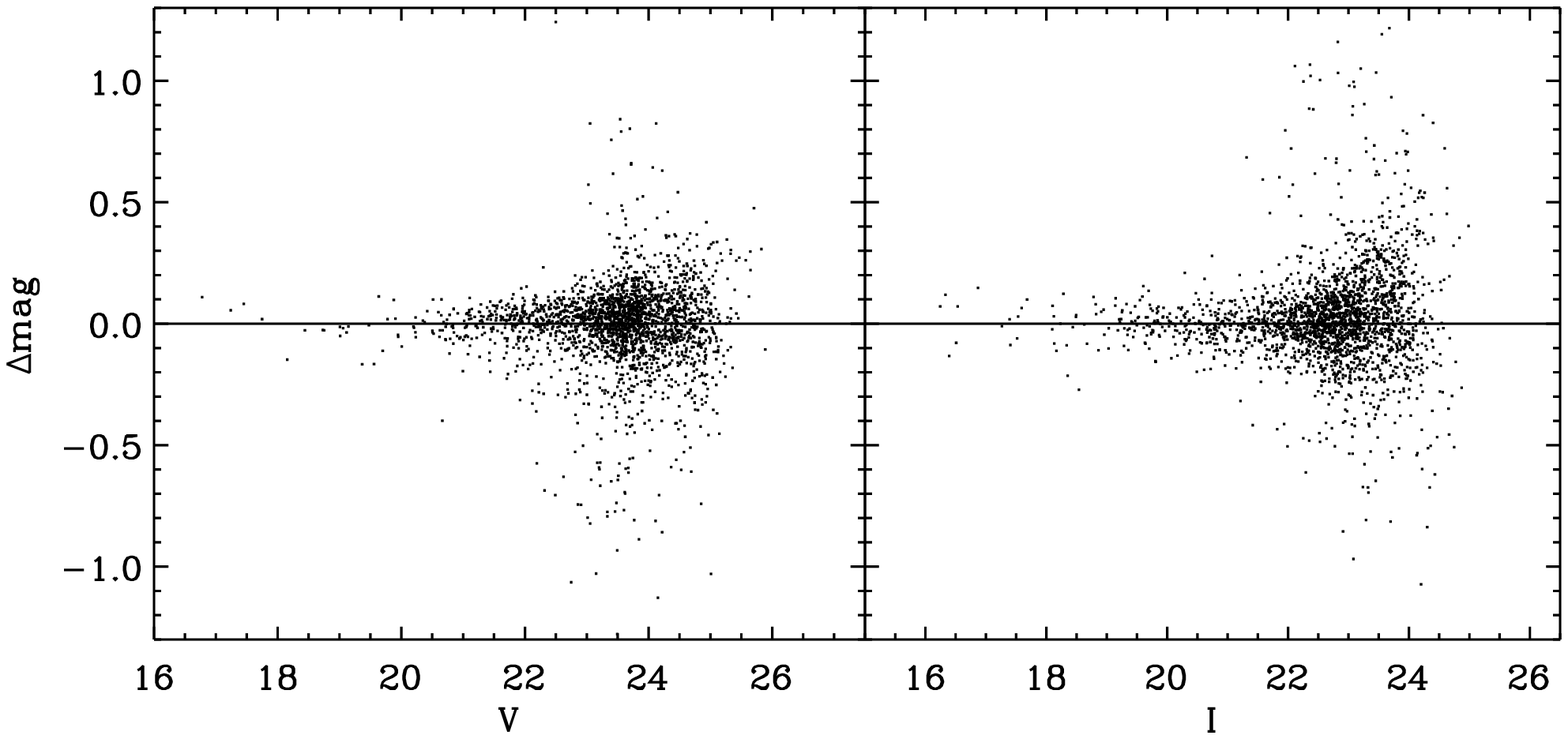}
\caption{Magnitude difference for stars in the overlapping regions between different pointings and with 
double measurements as a function of calibrated magnitudes (left: \mv\, band; right: \mi\, band). 
Here we plot only the stars that passed the selection criteria in sharpness and photometric errors (see Fig.~\ref{fig:sharp}). 
The scaled median absolute deviation at the bright end side of the distribution is 0.054 mag for the V-band ($19 < V < 21$) and  0.065 mag for 
the I-band ($18 < I < 20$), corresponding to an error in the individual measurement of 0.038 mag and 0.046 mag. 
At the faint end side these numbers are 0.14 and 0.13 ($24 < V < 25$ and $23 < I < 24$, respectively), corresponding 
to error in the individual measurement of  0.1 and 0.09 mag. At the bright end 
the photometric errors are 0.033 and 0.035 for \mv and  \mi band, respectively, and at the faint end they are 0.076 and 0.069 
at the faint end. Therefore, the magnitude differences from stars with multiple measurements are inflated 
by an extra 0.01 and 0.03 mag, and at the faint end by 0.06 and 0.06 mag. \label{fig:deltamag}}
\end{figure*}

The final step is to merge the catalogues for the various pointings in one catalogue containing 
a unique set of objects. This is done by performing weighted averages on the observed quantities 
for stars with multiple measurements.

At this stage we also evaluate which objects we consider as ``stars'' on the basis of their PSF fitting
shape parameters and errors in magnitudes. Figure~\ref{fig:sharp} shows the run of the 
DAOPHOT photometric parameters ``sharpness'' and ``chi'', as well as the overall magnitude 
error as a function of the calibrated magnitudes for the sample of unique (combined) measurements. 
We exclude a priori those objects with relatively large magnitude errors, i.e. 
$\sigma_V > 0.3$ mag and $\sigma_I > 0.4$ mag. We then examine the distribution of 
sharpness versus magnitude for this sub-sample of stars, deriving its scatter per 
bin of magnitudes. Given the asymmetric distribution around null values of the 
sharpness parameter, we retain as bona-fide stars those objects that fall within the region 
defined by the exponential envelopes fitted 
to $+3 \times$ and $-2 \times$ the scatter\footnote{The scatter 
is derived in an iterative manner and it is the scaled median absolute deviation (m.a.d.) 
of the sharpness per magnitude bin of 0.5mag.}. 
The catalogue of bona-fide stars consists of 19286 stars including the control field and 18919 in the main mosaic area 
(see Table~\ref{tab:catal}). 

Treating photon counts of each star as a Poisson variable, 
a S/N=10 would correspond to an error of magnitude of $\sim$0.1mag, and 
a S/N =5 to $\sim$0.2mag. This would imply for this data-set a S/N=10 at the magnitude $V_{\rm 10} \sim 24.8$,  $I_{\rm 10} \sim 23.6$, 
correspondig to 11737 stars in the main mosaic area. The magnitude limits 
at S/N =5 are $V_{\rm 5} \sim 25.4$,  $I_{\rm 5} \sim 24.0$. Here we have quoted the shallowest of the values among all fields and chips. 
The number of stars in the main mosaic area above the S/N=5 and S/N=10 magnitude limits is 15238 and 11737, respectively.

Figure~\ref{fig:deltamag} 
shows the magnitude differences of the bona-fide stars with double measurements. These differences will be used to estimate the 
typical magnitude and color error in various magnitude bins (see Sect.~\ref{sec:cmd}); in these values are therefore included 
internal photometric errors, errors from aperture corrections, errors due to the photometric calibration applied to different pointings, 
and effects of crowding (including possible mismatches of stars in the most crowded, inner regions). 

In order to understand in more detail the role of crowding in the error budget and 
on the completeness limit from the inner to the outer parts, we also perform artificial star tests on pointing 06. This is  
representative of the four central, most crowded pointings (see Fig.~\ref{fig:cov_img}). Since the FORS f.o.v. is 6.8\arcmin $\times$ 6.8\arcmin, 
one single pointing already extents from the center of Phoenix to about half nominal tidal radius (VDK91, MD99).

We add artificial stars to the combined V and I images for C1 and C2 separately, 
using the PSF-model derived for pointing 06. The coordinates and magnitudes are randomly chosen 
from a uniform distribution, and cover the whole extent of the pointing and instrumental magnitudes of stars in Phoenix 
(corresponding to calibrated magnitude in the ranges \mv [20,27] and \mi [18.5,26.5]). We inject 200 artificial stars 
per band per chip (about 5-10\% of the observed number of stars in these chips), and repeat the experiment 5 times. Finally we apply the same 
reduction and selection procedure that we used to produce the final catalogue of stars. We find that crowding does not 
significantly affect the 50\% completeness limit 
(i.e. the magnitude at which the fraction of recovered versus injected number of stars is 0.5): in \mi band this is found to be 
constant around \mi$\sim$23.5-23.8 over the explored 
range of R, while in \mv is \mv$\sim$24.2 in the inner 1.5\arcmin, and at larger R it remains constant at \mv$\sim$25. These values are 
very similar to the S/N$=10$ limits quoted above. 

The difference of recovered and input magnitudes of the artificial stars 
as a function of recovered magnitude is fairly symmetric around zero, showing that - except 
for a handful of stars - blending is not affecting the recovered magnitudes. 
The scaled m.a.d. between the difference of input and recovered calibrated magnitude is smaller than those determined using the 
stars from double measurements, e.g. in the most crowded region is 0.07~mag for \mv between 24,25 and 0.09~mag for \mv between 23,24. 
The errors from Fig.~\ref{fig:deltamag}, which we use throughtout the analysis, can then be considered as conservative.

\section{The overall structure of Phoenix} \label{sec:str}

\subsection{2D distribution} \label{sec:2D}
\begin{figure}
\begin{center}
\includegraphics[width=\linewidth]{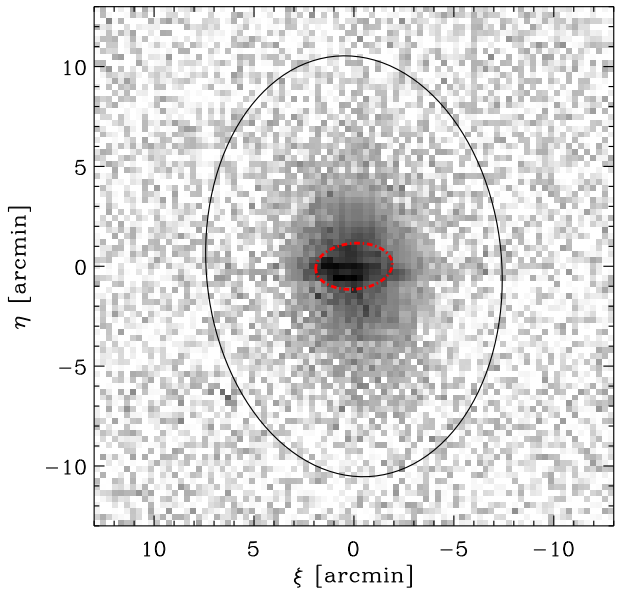}
\caption{Hess diagram in bins of 0.3\arcmin of the spatial distribution of the stars down to $V = 24.8$,  $I = 23.6$ (S/N$>$ 10). 
The solid ellipse shows the nominal tidal radius derived in this work, $r_t =  10.56' \pm  0.15'$ ($e=0.3$, P.A.$=$ 5\degr); for the 
inner dash-dotted ellipse see caption of Fig.~\ref{fig:cov_img}.
\label{fig:fov_hess}}
\end{center}
\end{figure}

\begin{figure}
\includegraphics[width=\linewidth]{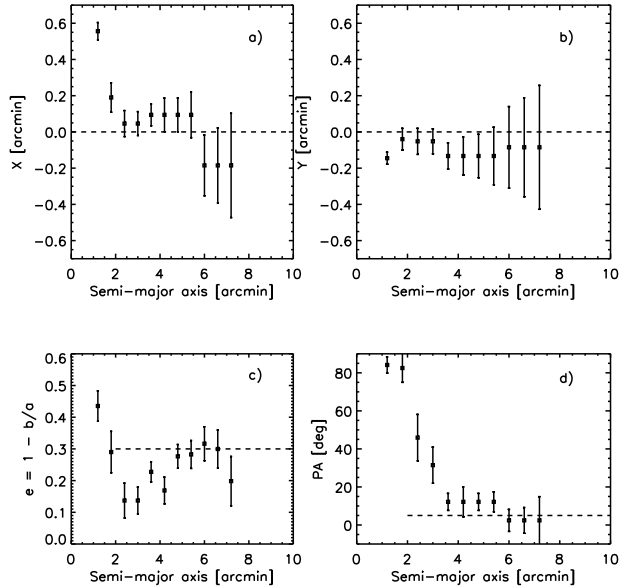}
\caption{The variation with projected radius of the central position ($x$ and $y$ coordinate, a) and b), respectively), 
c) ellipticity and d) position angle for the entire stellar population of Phoenix, for the stars brighter than 
$V = 24.8$,  $I = 23.6$ (S/N$>$ 10) from our FORS imaging data. 
The variation of 
the central position is with respect to the value listed in Mateo (1998).The horizontal lines in panel c) and d) 
indicate the values of the ellipticity and P.A. found for the outer Phoenix population by Martinez-Delgado et al. 1999.
\label{fig:par90}}
\end{figure}

When analyzing the 2D distribution of stars in Phoenix, 
MD99 found a sharp variation in the trend of the ellipticity ($e= 1 - b/a$, 
where $b$ and $a$ are the minor and major axes of the ellipse) and 
position angle (P.A.) of the isophleths with radius, with $e$ 
changing from 0.4 to 0.3 and the P.A. from 95\degr to 5\degr around 115\arcsec. 
The authors noted that the majority of young stars were contained in the 
flattened central ``component'' tilted with respect to the main body of the galaxy and 
the authors proposed this may be a disk-halo structure, as for large spiral galaxies.

We construct Hess diagrams of the spatial distribution of Phoenix stars, using the 
sample at S/N $>$ 5 and 10 (brighter than $V_{5} \sim 25.4$,  $I_{5} \sim 24.0$, and 
than $V_{\rm 10} \sim 24.8$,  $I_{\rm 10} \sim 23.6$, respectively), in bins of 0.3\arcmin. 
Since in some parts 
there appear to be voids or enhancements in the stellar number counts 
with respect to the surroundings, e.g. in the overlap region between pointings, we prefer 
not to smooth the spatial distribution because this may enhance such features. 
Figure~\ref{fig:fov_hess} shows the resulting distribution for the S/N$>$10 sample: 
it is clearly visible that the inner parts display a different distribution than 
the outer parts, being almost perpendicular to each other. 

We run the IRAF task ELLIPSE on the above Hess diagrams, letting the center, $e$ and P.A.
 of the isophotes at different radii 
as free parameters. The results are shown in Fig.~\ref{fig:par90} for the 
S/N$>$10 sample:  
within a radius of 1 and 2 arcmin, the center appears to be located 
about 0.6 -0.2 arcmin east and 0.1 arcmin south from the value listed in \citet{mateo1998};
 the ellipticity decreases from a value of 0.44$\pm$0.05 at 1\arcmin to 0.1 at 
about 3\arcmin, to increase again, with an average value of 0.28$\pm$0.03 at $R \ge $ 4.5\arcmin; 
the P.A. varies from about 80$^{\circ}$ within 2\arcmin and decreases reaching an average value 
of $8^{\circ} \pm 4^{\circ}$ at $R \ge $ 4.5\arcmin. These results are very similar to those 
resulting from the S/N$>$ 5 sample. They are also in good agreement 
with those of MD99, even though their results on the 2D distribution may have been more affected by crowding because of the 
larger seeing of the observations. We discuss the possibility of this being a disk-halo structure in 
Sect.~\ref{sec:variations}.

The intermediate values of ellipticity and P.A. that can be seen 
around 2\arcmin\,are probably due to the super-position of the inner and outer 
component. 

\subsection{Surface number density profile} \label{sec:surf}
\begin{table} 
\caption{Values of the density of contaminants [number arcmin$^{-2}$] as derived 
from the weighted mean of the outer points of the surface number profile of the 
various populations (column 2), from the objects at elliptical radii $ > $12.6\arcmin 
(column 3) and from pointing 17 (column 4). Note that since the MS display an asymmetric distribution we 
do not derive surface number profiles for them. For those populations that have stars as 
faint as the $V_{\rm 5}$, $I_{\rm 5}$ magnitude, we perform two determinations: one 
considering the stars brighter than the $V_{\rm 10}$,  $I_{\rm 10}$ limit (labelled as 10) and 
those brighter than the $V_{\rm 5}$, $I_{\rm 5}$ one (labelled as 5). Note that the majority of the determinations 
are compatible within 1$\sigma$ between the 3 methods. 
}
\label{tab:denscontam}
\centering
\begin{tabular}{lccc}
\hline
\hline
 & weighted mean & $R > $12.6\arcmin  & field 17\\
\hline
ms     & X                &  0.012        $\pm$     0.006 & 0.022 $\pm$0.022 \\
ms 10  & X                &  0.0                          & 0.022$\pm$0.022 \\
bl     & 0.11$\pm$ 0.02   &  0.092        $\pm$      0.017  & 0.087$\pm$0.04     \\
rc     & 0.543$\pm$  0.055&  0.474       $\pm$      0.038   & 0.281$\pm$0.078   \\
agbbump& 0.056$\pm$0.018 &              0.037       $\pm$      0.011  & 0.065$\pm$0.037    \\
rhb    & 0.338$\pm$  0.046&  0.361       $\pm$      0.033      & 0.238$\pm$0.072 \\
bhb  5 & 0.07$\pm$  0.023 &  0.092 $\pm$      0.017            & 0.043$\pm$0.031\\
bhb 10 & 0.053$\pm$  0.020 &  0.049$\pm$  0.012                &  0.022$\pm$0.022\\
rgb    & 0.055$\pm$  0.020&  0.046        $\pm$      0.012     & 0.022$\pm$0.022\\
all  5 & 8.44 $\pm$ 0.24  &  8.15      $\pm$       0.16        & 7.40$\pm$0.40\\
all 10 & 5.52 $\pm$ 0.19 &  5.35$\pm$0.13                      & 5.58$\pm$0.35\\
\hline
\end{tabular}
\end{table}

There have been previous determinations of the surface number density profile of Phoenix, 
with results rather different from each others: VDK91 used a scanned 
ESO/SRC IIIaL plate and found that the surface brightness profile is not well fit by a King profile 
but decreses exponentially reaching 
the zero level at 8.7\arcmin\,from the center; 
on the other hand, MD99 find that the profile 
can be well fit by a King profile, and determine a nominal tidal radius of $r_t = 15.8_{-2.8}^{+4.3}$ arcmin. 

We derive the 
surface number density profile of the overall stellar population of 
Phoenix using both the data-sets with S/N$>$ 5 and 10. We sample the number counts in 
elliptical bins\footnote{Throughout the manuscript the projected elliptical radius 
of a point (x, y) is defined as $R= \sqrt{x^2+ y^2/( 1 - e )^2}$, where $e$ is the considered ellipticity, 
and the galaxy is assumed to be centred on the origin, with its major axis aligned with the x-axis.}. For the 
values of ellipticity of the elliptical bins we adopt the values for the outer components of MD99, i.e. 
$e = 0.3$ and P.A.$=$ 5\degr, given the good agreement between our results 
and those of MD99, and the fact that MD99 derived those parameters from a single plate rather than a mosaic of images. 

Given the discrepant measurements of the extent of Phoenix in the literature, we decided to estimate empirically the density of contaminant objects (MW foreground stars and unresolved galaxies), rather than relying on uncertain selection of a region supposedly free of Phoenix stars. 
The following two approaches were adopted: a) we compute the number density profile in elliptical annuli whose semi-major axis is $\la$13\arcmin, 
i.e. the whole ellipse is contained in our main mosaic area; we analyze the outer parts of the profile, and 
adopt as density of contaminants a weighted average of the outer points, where the profile becomes approximately 
flat; in this case this is the outermost 3\arcmin; 
b) we analyze the number density profile along the projected minor axis of the galaxy; this is because 
in that direction our mosaic covers a region corresponding to rather large elliptical radii, 18.6\arcmin, i.e. a region 
beyond the largest literature measurement of the nominal tidal radius of the galaxy (15.8\arcmin, MD99); 
along this axis, we find that the profile flattens at $\sim$8.8\arcmin, corresponding to an elliptical radius 
of 12.6$\arcmin$; we therefore expect the great majority of the 
objects found at elliptical radii larger than 12.6\arcmin\,to be 
composed by contaminants; we use then the objects located at elliptical radius $>$12.6\arcmin\,to determine the density of contaminants. 
The values we obtain with the two methods are consistent with each other (see Table~\ref{tab:denscontam}). As 
a consistency check we compare these determinations to the ones 
derived from pointing 17 (see Table~\ref{tab:denscontam}), i.e. the pointing displaced from the 
rest of the mosaic area, and also find good agreement; however, given the small area 
covered by pointing 17, we will be using the other two methods in the following analysis.

\begin{figure}
\includegraphics[width=\linewidth]{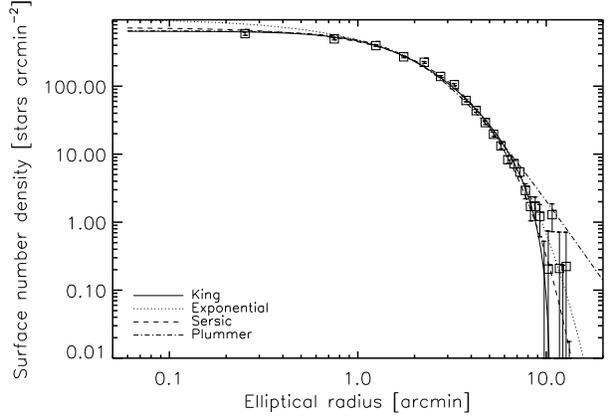}  
\caption{The surface number density profile for the Phoenix dwarf galaxy with overlaid best-fitting King, Sersic, 
exponential and Plummer models 
(solid, dashed, dotted, dash-dot lines, respectively). The density of contaminants, $5.52\pm0.19$ stars arcmin$^{-2}$ 
calculated from a weighted average of the outer 3\arcmin, has been subtracted from each point. 
The error bars are obtained by summing in quadrature the errors from Poisson statistics and the error in 
contamination. The central point was excluded from the fit. Our best fit to the data is a Sersic profile with Sersic radius, 
$R_{\rm S}= 1.82' \pm 0.06'$ and m$=$ 0.83$\pm$0.03 with a reduced $\chi^2 = 1.8$; a King profile 
with core radius $r_c = 1.79' \pm  0.04'$ and tidal radius $r_t =  10.56' \pm  0.15'$ gives a reduced 
 $\chi^2 = 3.5$;. 
The best-fitting parameters for the different models are summarized in Table~\ref{tab:par_oldint}.
\label{fig:surfbr}. }
\end{figure}

We subtract the density of contaminants derived from methods a) and b) from the 
number density profile of Phoenix sampled in bins of 0.5\arcmin\,and we fit  
this contamination-subtracted profile   
to several surface brightness models using a least-square fit to the data. 
We used an empirical King profile \citep{king1962}, 
an exponential profile, 
a Sersic profile \citep{sersic1968} and a Plummer model \citep{plummer1911}. 

\begin{table*} 
\caption{Parameters of best-fitting King model (core radius, $r_{\rm c}$, tidal radius, $r_{\rm t}$), 
Sersic model (Sersic radius, $R_{\rm S}$, shape parameter, $m$), exponential model (scale radius, $r_{\rm Ex,h}$) 
and Plummer model (scale radius, $b$) 
for various stellar populations in Phoenix dSph. For each model we list the corresponding reduced $\chi^2$, 
 $\chi_{\rm red}^2 = \chi^2/(N - \nu)$, where N is the number of points in the surface number count profile and 
$\nu$ the number of degrees of freedom. The 
errors in the parameters are from Montecarlo simulations but agree well with the
 formal errors from the fitting procedure. These values of the parameters were derived by using the 
density of contaminants from the weighted average of the last points of the surface number profile. 
The best-fitting profiles have good values of the reduced $\chi^2$.
}
\label{tab:par_oldint}
\centering
\begin{tabular}{lcccccccccc}
\hline
\hline
\multicolumn{1}{l}{{}} &  \multicolumn{3}{c}{{King}} & \multicolumn{3}{c}{{Sersic}} &
\multicolumn{2}{c}{{Exponential}} & \multicolumn{2}{c}{{Plummer}} \\
\hline
  & $r_{\rm c}[']$  & $r_{\rm t}[']$ & $\chi_{\rm red}^2$& $R_{\rm S}[']$  & $m$ & $\chi_{\rm red}^2$& $r_{\rm Ex,h}[']$ & $\chi_{\rm red}^2$& $b[']$ & $\chi_{\rm red}^2$\\
\hline
All  5         &   1.82$\pm$0.04 & 10.77$\pm$0.17 & 3.0 & 1.76$\pm$0.07 & 0.86$\pm$0.03 & 2.0 & 1.40$\pm$0.01 & 2.9 & 2.50$\pm$0.02 & 6.7 \\
All 10         &   1.79$\pm$0.04 & 10.56$\pm$0.15 & 3.5 & 1.82$\pm$0.06 & 0.83$\pm$0.03 & 1.8 & 1.37$\pm$0.01 & 3.2 & 2.43$\pm$0.02 & 7.9\\
\hline
BL             &   0.73$\pm$0.13 & 9.53$\pm$0.70  & 1.0  & 0.53$\pm$0.26 & 1.25$\pm$0.24 & 0.9 & 0.86$\pm$0.05 & 0.9 & 1.47$\pm$0.09 & 0.9\\
RC (23.2-23.4) &   1.11$\pm$0.09 & 9.98$\pm$0.32  & 1.9  & 1.45$\pm$0.27 & 0.82$\pm$0.11 & 1.1 & 1.05$\pm$0.04 & 1.2 & 1.79$\pm$0.06 & 1.9  \\
RC (23.4-23.6) &   1.37$\pm$0.08 & 10.40$\pm$0.25 & 2.4  & 2.10$\pm$0.26 & 0.66$\pm$0.08 & 0.8 & 1.17$\pm$0.03 & 1.5 & 2.00$\pm$0.06 & 3.0 \\
RC (23.6-23.8) &   1.91$\pm$0.09 & 10.66$\pm$0.23 & 2.0  & 2.54$\pm$0.23 & 0.62$\pm$0.07 & 1.1 & 1.37$\pm$0.03 & 2.1 & 2.36$\pm$0.06 & 3.7 \\
RGB bump       &   1.81$\pm$0.13 & 11.08$\pm$0.38 & 0.9  & 1.88$\pm$0.24 & 0.83$\pm$0.08 & 0.9 & 1.39$\pm$0.04 & 1.0 & 2.34$\pm$0.06 & 2.2 \\
RHB            &   2.40$\pm$0.13 & 10.42$\pm$0.18 & 2.2  & 2.85$\pm$0.19 & 0.61$\pm$0.04 & 1.9 & 1.48$\pm$0.03 & 3.5 & 2.57$\pm$0.04 & 5.8 \\
BHB 10         &   2.68$\pm$0.33 & 10.72$\pm$0.61 & 1.3  & 2.87$\pm$0.47 & 0.65$\pm$0.10 & 1.2 & 1.63$\pm$0.07 & 1.5 & 2.89$\pm$0.15 & 1.8 \\
BHB  5         &   2.15$\pm$0.30 & 12.05$\pm$0.75 & 0.7  & 1.91$\pm$0.36 & 0.91$\pm$0.08 & 0.9 & 1.61$\pm$0.06 & 0.9 & 2.70$\pm$0.12 & 1.8 \\
RC             &   1.77$\pm$0.04 & 10.28$\pm$0.12 & 4.6  & 2.24$\pm$0.11 & 0.68$\pm$0.03 & 1.6 & 1.31$\pm$0.02 & 4.2 & 2.28$\pm$0.03 & 9.1 \\
RGB            &   1.75$\pm$0.07 & 10.66$\pm$0.18 & 2.3  & 2.18$\pm$0.19 & 0.71$\pm$0.05 & 0.7 & 1.33$\pm$0.03 & 1.7 & 2.25$\pm$0.05 & 4.6 \\
\hline
\end{tabular}
\end{table*}

The King model has been extensively used to describe the 
surface density profile of dSphs.
\begin{equation}
I_{\rm K}(R)= I_{\rm 0,K} \ \left( \frac{1}{\sqrt{1 + \left(\frac{R}{r_c}\right)^2}} - 
\frac{1}{\sqrt{1 + \left(\frac{r_t}{r_c}\right)^2}} \right)^2
\end{equation}
It is defined by 3 
parameters: a characteristic surface density, $I_{\rm 0,K}$, core radius, $r_{\rm c}$ 
and tidal truncation radius, $r_{\rm t}$.

We note that excesses of stars have been found beyond the 
King tidal radius in Fornax \citep[e.g.][]{coleman2005, battaglia2006}, Sculptor 
\citep[e.g.][]{coleman2005scl, battaglia2008b} and in a number of other LG 
dSphs, e.g. Carina \citep{majewski2005}, Draco \citep[e.g.][]{wilkinson2004}, Ursa Minor \citep[e.g.][]{IH1995, MD2001}. 
Since the tidal radius is set by the tidal field of the 
host galaxy, the observed excesses of stars have been interpreted as tidally stripped stars.
However, if 
dwarf galaxies are embedded in massive dark matter haloes, this tidal 
radius loses its  
meaning of tidal truncation radius (we will therefore refer to it as ``nominal tidal radius'') and 
such excesses of stars need not be tidally stripped stars but 
can simply due to the King model not providing the best representation of the surface number count profiles of dSphs at large radii.

The Sersic profile is known to provide a good empirical formula to fit the projected light distribution of 
elliptical galaxies and 
the bulges of spiral galaxies \citep[e.g.][]{caon1993, caldwell1999, graham2003, trujillo2004} 
and also provides a good representation of the number surface density of some dSphs \citep[e.g.][]{battaglia2006, battaglia2008b}: 
\begin{equation}
I_{\rm S}(R)= I_{\rm 0,S} \ exp \left[ - \left(\frac{R}{R_{\rm S}}\right)^{1/m}\right]
\end{equation}
where $I_{\rm 0,S}$ is a scale surface density, $R_{\rm S}$ is a scale radius and $m$ is the 
surface density profile shape parameter. 

We find that the best-fitting parameters are very similar when using the contaminant density derived 
from either method a) or b). In the following we will adopt method a) because the fits 
to the surface brightness profile produce slightly better $\chi^2$ values; method a) will be 
replaced by method b) only when analyzing the young main sequence stars because of 
their asymmetric distribution (see Sect.~\ref{sec:spatdistr}). 

The results of the fit for the S/N$>$10 sample are shown in Fig.~\ref{fig:surfbr} and summarized 
in Table~\ref{tab:par_oldint}. In our analysis, the Sersic profile is the one that fits the data best, yielding 
a reduced $\chi^2 = 1.8$ for a scale radius $R_S = 1.82' \pm 0.06$\arcmin\,and $m= 0.83 \pm 0.03$. 
An exponential profile with exponential radius 1.37\arcmin$\pm$0.01\arcmin (reduced  $\chi^2 = 3.2$) and 
a King profile with core radius $r_c = 1.79' \pm 0.04$\arcmin, tidal radius $r_t =  10.56' \pm  0.15$\arcmin\,(reduced  $\chi^2 = 3.5$) 
also reproduce the data rather well; the Plummer profile instead clearly 
overpredicts the surface number density at R$\ga$7\arcmin. The results from the 
S/N$>$5 sample are very similar, both in terms of shape parameters and best-fitting profiles. 
Because 
of the intrinsic uncertainty in determination of crowding corrections and 
because we are mostly interested in the large scale properties of Phoenix, we are not applying a crowding correction factor. 
Instead, we have performed the fit excluding the central point of the surface number count profile. It was anyway shown in Sect.~\ref{sec:cal} 
that the effects of crowding are not significant. This is confirmed by the results of the 
fit to the surface number count profile derived from the sample of stars brighter than the 50\% completeness limit at $R<$1.5\arcmin: 
the best-fitting parameters are in very good agreement 
with those derived from the sample S/N$>$10, and they yield very similar $\chi^2$ values.

In order to assess whether Sersic and King profiles perform better because of one more free parameter 
when compared to exponential an Plummer profiles, we check how the performance of 
the best-fitting profiles would be ranked using the Akaike information criterion \citep{akaike1973} 
in the form $\chi^2 + 2 \times k$, where $k$ is the number of free parameters in the fit: the ranking of the best-fitting 
profiles remains the same as according to the reduced $\chi^2$ values, therefore hereafter we consider only the 
reduced $\chi^2$ values.

Our determination of the tidal radius $r_t =  10.56' \pm  0.15$\arcmin\,is smaller than the one from MD99, 
$r_t = 15.8_{-2.8}^{+4.3}$ arcmin; 
note that the value from MD99 was derived using the surface number density 
between projected radii of approximately 2\arcmin\,and 5\arcmin, and extrapolating the behaviour at 
larger distances, while our determinations comes from a much wider area.

Since the outer parts of the object are almost equally well fit by a 
range of profiles, continuous and truncated, it is therefore difficult to establish whether 
Phoenix was tidally truncated by the Milky Way or not.

In the following we will use the King tidal radius we obtained as an indication 
of the extent of Phoenix, and we will refer to it as 
the ``nominal'' tidal radius as it is unclear whether this is the radius of 
tidal truncation of the galaxy. 

\begin{figure*}
\includegraphics[width=0.45\linewidth]{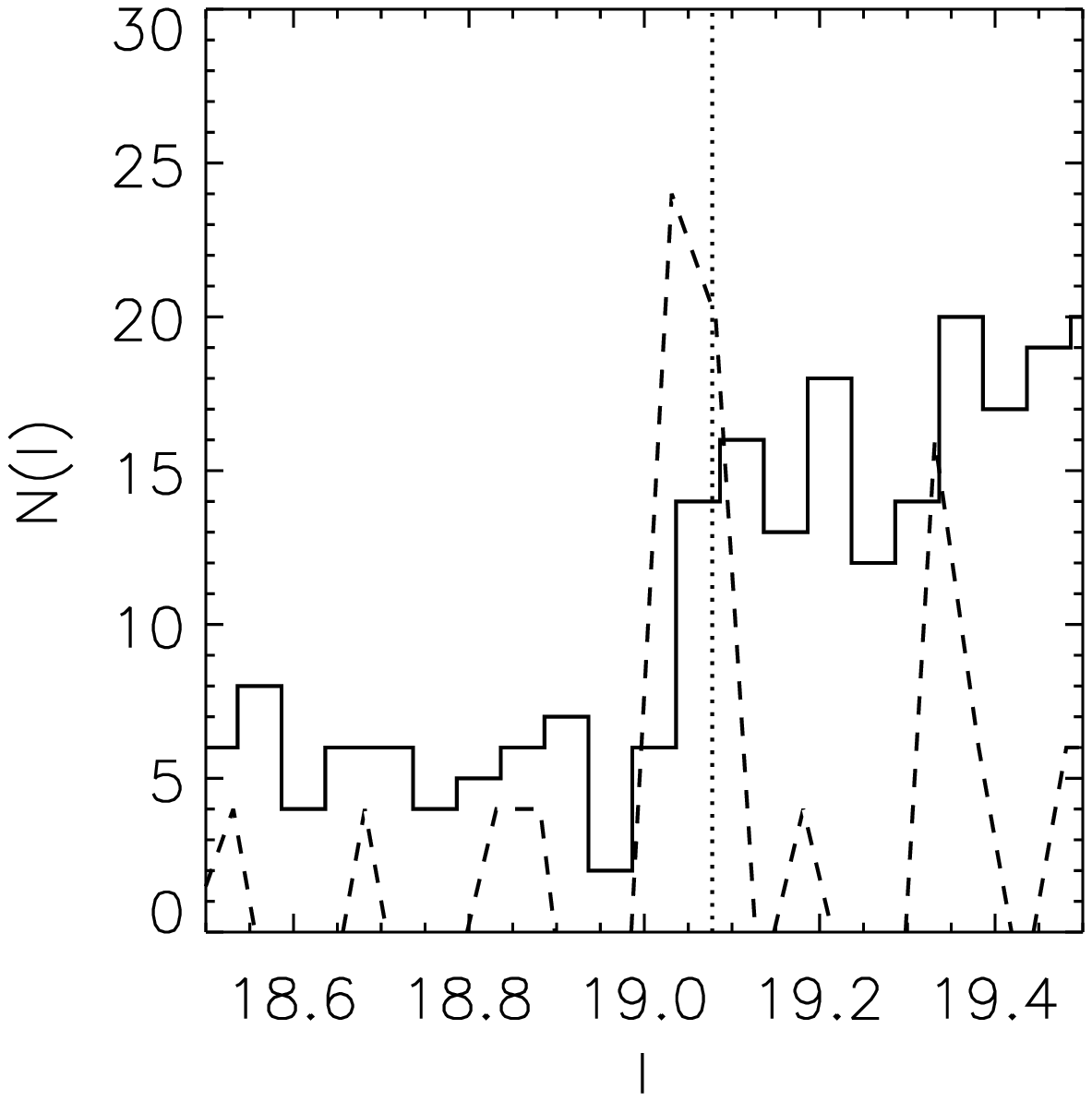}  
\includegraphics[width=0.45\linewidth]{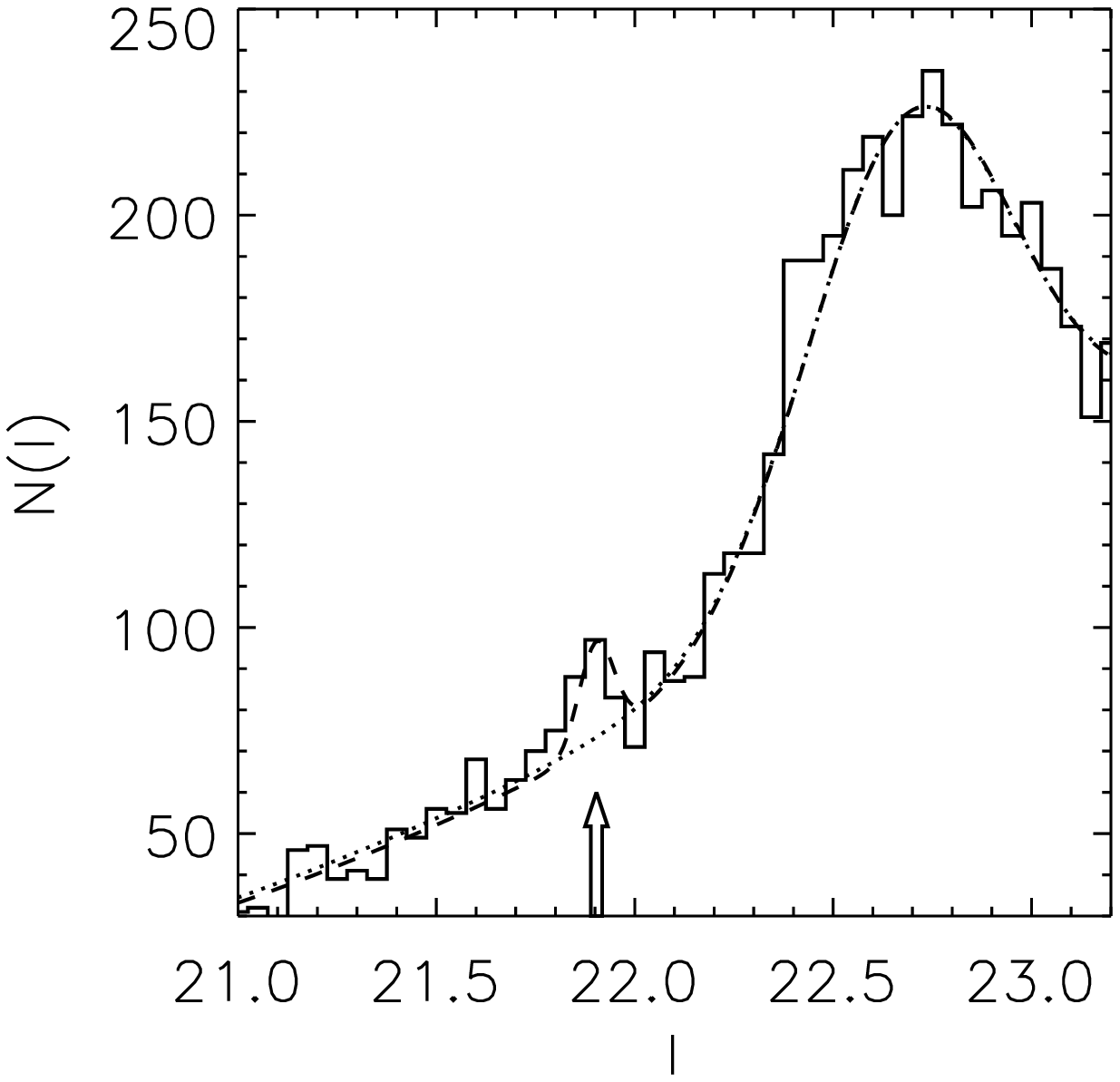}  
\caption{Luminosity function of Phoenix stars (solid histogram) in the magnitude range 18.5 $<$ \mi $<$ 19.5 (left) and 21.0 $<$ \mi $<$ 23.2 (right). 
Left: The vertical dotted line 
indicates the $I_{tip, RGB}$ found over about 100 random realizations of the luminosity function convolved with 
a Sobel filter of kernel [-2,0,2] over different binnings (see the text for more details); the dashed line shows one of these random realizations. 
Right: the arrow indicates the position of the RGB bump (see text); the dashed line shows the fit of Eq.~(4) to the histogram, while the dotted line 
is the model LF excluding the Gaussian representing the RGB bump.  
\label{fig:lf_tip}}
\end{figure*}

\section{Tip of the RGB} \label{sec:tip}
We derive the tip of the RGB of Phoenix in \mi, $I_{\rm RGBT}$, and 
compare it to the determinations from previous works. In addition to providing an independent distance determination, 
this also offers a possibility to check our photometric calibration. 

For this, we convolve the luminosity 
function (LF) in the \mi\, band for stars with 18 $<$ \mi $<$ 21 with a Sobel edge-detection kernel [-2,0,2] \citep{lee1993}, 
which gives a maximum where the discontinuity in the luminosity function is greatest (see Fig.~\ref{fig:lf_tip}, left). 

In order to factor in the uncertainties due to the binning of the LF, we perform the determination over 100 
random realizations of the luminosity range used for the calculation - that is, we allow the brightest end of the 
range to vary between \mi\, 18 and 19, and the faintest end between 21 and 22. For each of these random realizations, we 
perform the convolution using bin sizes decreasing from 0.15 mag to 0.05 mag with step of 0.01 mag. 
Since there are several discontinuities in the LF due to various features in the colour-magnitude diagram (see below), 
we restrict our search for the maximum 
to the region 18 $<$ \mi $<$ 20; this choice comes from an educated guess from a visual 
inspection of the CMD, and from literature values of $I_{\rm RGBT}$.

We find $I_{\rm RGBT}= 19.08\pm0.06$, where the error is the scaled m.a.d. of the distribution of values. 
This agrees with previous measurements 
such as those by H09, who find $I_{\rm RGBT}= 19.14\pm0.02$, and MD99   
$I_{\rm RGBT}= 19.00\pm0.07$ mag. 
These authors use a slightly different version of the Sobel filter, i.e. with a kernel [1,2,0,-2,-1], 
but we checked that the use of this 
kernel makes no difference to our determination. 

We adopt a reddening value E(B-V)$=0.016$  (see Table~\ref{tab:par}), 
derived from the reddening maps of our Galaxy 
available at http://irsa.ipac.caltech.edu/applications/DUST/ centered on the coordinates of the 
Phoenix dwarf galaxy. This corresponds to an extinction in 
\mv\, $A_V = 0.05$ mag and in \mi band $A_I = 0.024$ \citep{cardelli1989}, 
yielding a dereddened magnitude for the RGB Tip $I_{\rm RGBT0}\sim 19.06$. 

As shown in Lee, Freedman \& Madore (1993) the absolute magnitude of the tip of the RGB in I-band changes 
by less than 0.1 mag around the value M$_I = -4.0$ 
in the metallicity range -2.2$<$[Fe/H]$<-0.7$, which includes the metallicity 
range expected for stars in Phoenix (see H09). Therefore by adopting an absolute magnitude M$_{I, RGBT} = -4.0$ and 
an extinction A$_I = 0.024$, a $I_{\rm RGBT}= 19.08$ would give a distance modulus (m-M)$_0 = 23.06 \pm 0.12$;  
the error is due to the combination of the estimated error on the determination of $I_{\rm RGBT}$ and the 
fact that we are neglecting the (weak) dependence of the absolute magnitude of the RGB tip in I-band with 
[Fe/H] (i.e. the 0.1mag variation of the magnitude of the tip around the value M$_I = -4.0$ mentioned above). 
This is in good agreement with the value from H09.  The distance inferred from this distance modulus is 409$\pm$23 kpc.

If we take explicitly into account the dependence of M$_{\rm I, RGBT}$  on global 
metallicity [M/H], using the calibration from \citet{bellazzini2004}, we would obtain a 
distance modulus  (m-M)$_0 =$23.07 and 23.12 for [$\alpha$/Fe] $= 0.0$ and $= +0.4$, respectively, 
fully compatible with the determination above. 
Here we made use of the formula from \citet{salaris1993}, 
$\mathrm{[M/H]} = \mathrm{[Fe/H]} + \mathrm{alog10}(0.638 \times 10^{[\alpha/\mathrm{Fe}]} + 0.362)$ and 
from \citet{dacosta1990}, $\mathrm{[Fe/H]} = -15.16 + 17.0 \times \mathrm{(V-I)}_{0,-3} -4.9 \times \mathrm{(V-I)}_{0,-3}^2$, 
where $\mathrm{(V-I)}_{0,-3}$ is the mean color of the RGB at absolute I mag $=-3$ that for our data is 
$\mathrm{(V-I)}_{0,-3} = 1.21$. 

\section{Stellar populations in Phoenix and their spatial variations} \label{sec:variations}
\begin{figure*}
\includegraphics[width=0.45\linewidth]{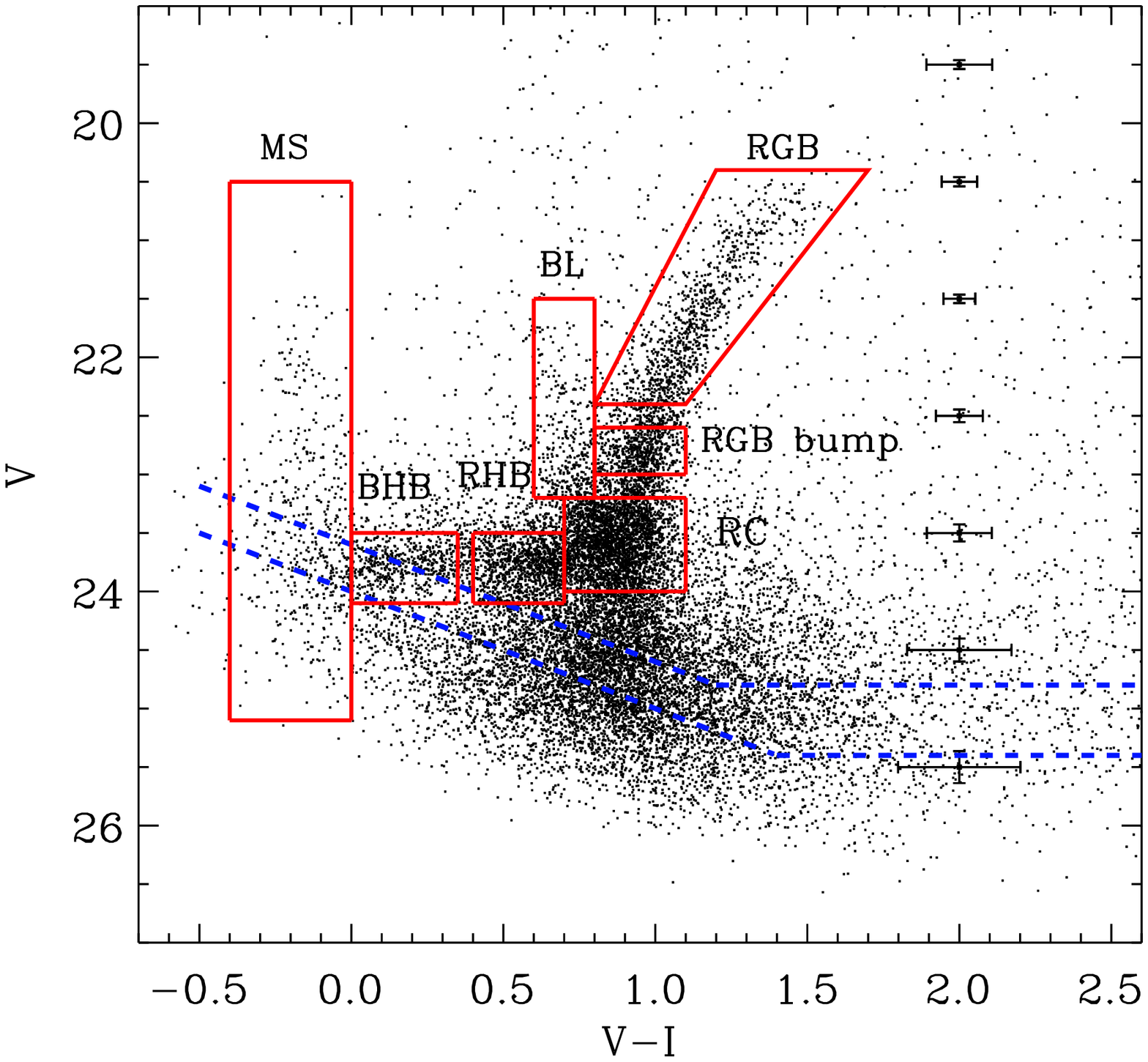}  
\includegraphics[width=0.45\linewidth]{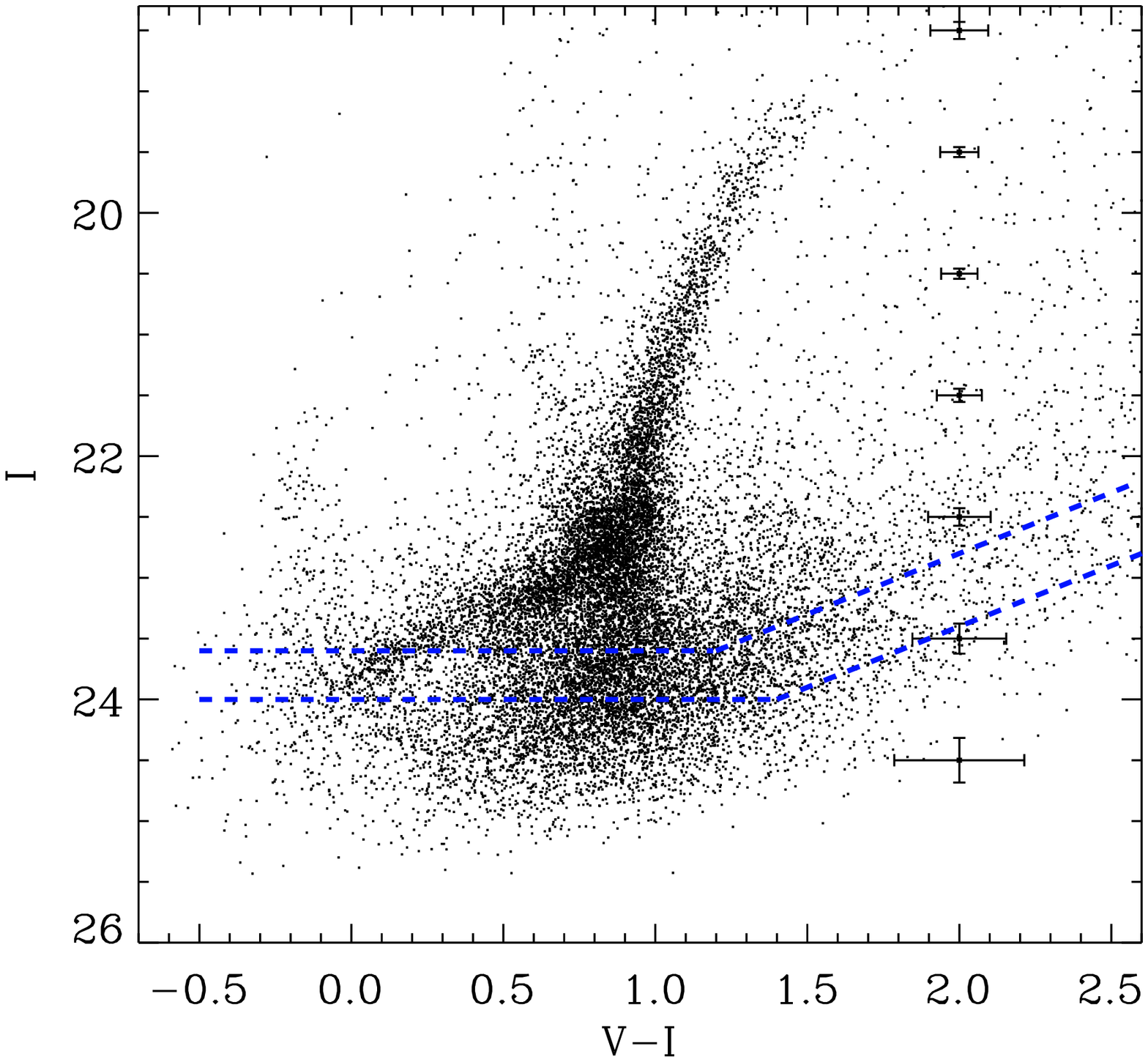}  
\caption{Colour-magnitude diagram for objects classified as stellar in the Phoenix dwarf galaxy 
(left: \mv\, vs \vi\,; right: \mi\, vs \vi). Here we do not include pointing 17, which was 
observed at a displaced location from the galaxy for checks on the foreground/background density. 
The dashed lines indicate the levels of S/N= 5 and 10 in both bands. The 
error-bars on the right indicate the typical magnitude and color error per magnitude bins of 0.1 mag, 
and they are derived from the magnitude differences of the stars with double measurements (see Fig.~\ref{fig:deltamag}). 
In the left panel we show the selection boxes for stars in different evolutionary phases, as 
indicated by the labels (see text).
\label{fig:cmd}}
\end{figure*}

\subsection{Stellar populations in the Colour-Magnitude Diagram} \label{sec:cmd}
Figure~\ref{fig:cmd} shows the resulting colour-magnitude diagram (CMD) for our FORS 
mosaic of the Phoenix dwarf galaxy (left: \mv band; right: \mi band; lines of 
$V_{\rm 10} \sim 24.8$,  $I_{\rm 10} \sim 23.6$ and $V_{5} \sim 25.4$,  $I_{5} \sim 24.0$ are plotted in the figure). 
The main features of the stellar population of Phoenix are clearly visible and indicated with boxes 
on the CMD: a well defined 
RGB, which should also contain a component of AGB stars; 
a clear detection of an horizontal branch, divided in a red (RHB) and blue (BHB) part, 
extending to \vi $\sim$ 0;  
a well populated red clump (RC); an approximately 
vertical sequence of young main sequence (MS) stars, centered at \vi $\sim -0.2$; a vertical 
sequence of stars emerging from the RC, at \vi $\sim$0.7, containing blue-loop (BL) stars. 

We refer the reader to \citet{holtzman2000} for a detailed discussion of the various features. Since we 
will use stars in different evolutionary stages to analyze how the stellar population mix varies 
throughout the galaxy, here we focus on how the features present in Phoenix CMD can 
be used as indicators of 
different age ranges. In evaluating 
the age range dominant for stars in a certain evolutionary phase we can use the 
information on the Phoenix SFH and chemical enrichment history derived by H09, i.e. that 
stars older than 6 Gyr have Z between 0.0002 and 0.0004, 
while the stars younger than 2 Gyr have Z between 0.001 and 0.002. 

From stellar evolutionary models is known that HB will contain stars predominantly $>$10 Gyr old (ancient), 
while the RGB stars will sample the whole stellar population mix, with the exception of the 
stars younger than about 1 Gyr. The younger end of the age distribution can be explored using the BL and MS stars: 
the MS stars above the $V_{\rm 10}$ , $I_{\rm 10}$ limit, selected as in \ref{fig:cmd}, are consistent with 
ages between 0.1-0.5 Gyr, while the selected BL stars are sampling slightly older stars, mainly 0.5-1 Gyr old. 

The RC contains 1-10 Gyr old stars, in a proportion changing with the SFH and metallicity of the 
stellar population. Using the predictions on the magnitude of RC stars from \citet{girardi2001} one can 
see that for the rather narrow and metal-poor metallicity range of stars in Phoenix, 
both the V and I magnitude can act as age indicators. 
Therefore we split the RC stars in magnitude bins of V $=$ 23.2-23.4, 23.4-23.6, 23.6-23.8 
as indicators of age ranges 2-5 Gyr, 5-8 Gyr and 8-12 Gyr, respectively. 
The mean magnitude of the RC (derived in Sect.~\ref{sec:bump}) indicates that this feature is dominated by 5-8 Gyr old stars.

\subsubsection{Bump on the RGB} \label{sec:bump}
A bump is visible along the RGB and above the RC, both in the CMDs in Fig.~\ref{fig:cmd} and in the LF (see Fig.~\ref{fig:lf_tip}, right). 
In order to understand whether this bump can be 
classified as an AGB or RGB bump, we derive its mean V and I magnitudes and compare them to 
the predictions from stellar evolutionary models. 

The I-band LF locally around the RC can be well fit with a combination of a polynomial (for the RGB) and  
a Gaussian function for the RC stars \citep{stanek1998}. To this function we add 
another Gaussian in order to fit the bump present at brighter magnitude than the RC:
\begin{eqnarray}
N(I)  & = & a + b \times I + c \times I^2 + d \times e^{-\frac{(I-I_{\rm RC})^2}{2 \sigma_{\rm RC}}}+ \nonumber \\
 &  & g \times e^{-\frac{(I-I_{\rm bump})^2}{2 \sigma_{\rm bump}}}, \nonumber \\ 
\end{eqnarray}
where $I_{\rm RC}$, $I_{\rm bump}$ are the mean magnitude of the RC and bump, respectively, and $\sigma_{\rm RC}$, $\sigma_{\rm bump}$ 
the dispersion. 

We fit Eq.~(4) to the LF between 19 $<$ \mi $<$ 23.5 in bins of 0.07 mag in  
a range of color broadly covering the RC, i.e. 0.7 - 1.2 (see Fig.~\ref{fig:lf_tip}, right). This gives 
$I_{\rm RC} = 22.7$ and $\sigma_{\rm I, RC}$ = 0.26 mag for the RC and $I_{\rm bump} = 21.90$ 
with $\sigma_{\rm I, bump} = $ 0.044$\pm$0.02 mag for the bump. 
We also fit a similar function to the LF in V band over the same colour range and 20 $<$ \mv $<$ 24.5; 
overall, such function gives a good representation of the data and yields 
$V_{\rm RC} = 23.54$ with a dispersion of $\sigma_{\rm V, RC} =$ 0.22 mag, and 
$V_{\rm bump} = 22.82$ with $\sigma_{\rm V, bump} = $ 0.13 mag. However, since the bump feature does not 
seem particularly well described by a Gaussian in the \mv band, we also derive the $V_{\rm bump}$ from 
the observed $I_{\rm bump}$ by fitting the color distribution of 
the stars in the range 0.8 $<$ \vi $<$ 1.1 with a Gaussian. This yields a $(V-I)_{\rm bump} = 0.95$, resulting in 
$V_{\rm bump} = 22.85$, in good agreement with our other determination. 

Using the distance modulus, the extinction and reddening values in Table~\ref{tab:par}, these 
would result in absolute magnitudes $M_{\rm V, RC} = +0.43$,$M_{\rm I, RC} = -0.38$, $M_{\rm V, bump} = -0.26$ 
and $M_{\rm I, bump} = -1.18$. The de-reddened color is  $(V-I)_{\rm bump, 0} = 0.92$

We examine the predicted dependency of the RGB and AGB bump magnitude and color as a function of age and metallicity, 
over the range 1 $\le$ age [Gyr] $\le$ 12 and 0.0001 $\le$ Z $\le$ 0.02, in 
Fig.~\ref{fig:bump} (top and bottom panels, respectively). For this, we use Padua isochrones \citep{girardi2000, marigo2008}, 
in which the onset and end of the various stellar evolutionary phases are conveniently indicated\footnote{http://stev.oapd.inaf.it/cgi-bin/cmd webpage}. 
We remind the reader that the SFH and chemical enrichment history derived by H09 for Phoenix 
show that the stars older than 6 Gyr have Z between 0.0002 and 0.0004, 
while the stars younger than 2 Gyr old have Z between 0.001 and 0.002. Using these isochrones, we find that the 
magnitudes of the detected feature are not 
compatible with those of an AGB bump for the range of metallicities of stars in Phoenix and are instead fully 
compatible the magnitudes and color of a RGB bump 
for stars with metallicity Z between 0.001 and 0.0001 (see Fig.~\ref{fig:bump}). 

Since the work of H09 was carried out using Basti isochrones \citep{pietrinferni2004, pietrinferni2006} 
and given that different sets of isochrones do not 
always predict similar magnitudes/colors for stars in the same evolutionary phase \citep[e.g.][]{gallart2005}, we 
verified if the same conclusion would be reached using Basti isochrones. In this case, there is still very good agreement 
between the 
observed bump magnitudes and colors and those predicted for an RGB bump in the metallicity range appropriate for 
Phoenix stars, but there is also a marginal consistency with the values predicted for an AGB bump. 
We then examine the LFs derived for single stellar populations of ages 3-5-8-10 Gyr and Z=0.0001-0.001-0.01 using the online tool on the 
Basti website, and we see that the RGB bump is always more populated than the AGB bump for the metallicities of Phoenix stars. 
Since in our data we are able to detect only 1 such clump, then this is most likely to be the most populated of the two, 
i.e. the RGB bump. Overall, then, the two sets of isochrones provide the same conclusion.

The RGB bump stars sample the intermediate and old age populations. For the same age and metallicities the AGB bump is expected at similar color, 
but about 0.5 mag brighter in the V-band than the RGB bump; within our dataset, however, we do not detect any evident feature at that magnitude. 

RGB bumps have been detected in many LG dwarf galaxies, both in dSphs, dIrrs and dTs \citep[see][and references therein]{monelli2010}; this is 
the first detection of the RGB bump in the Phoenix dwarf galaxy.

\subsection{Spatial distribution of stellar populations} \label{sec:spatdistr}
\begin{figure*}
\includegraphics[width=0.7\linewidth]{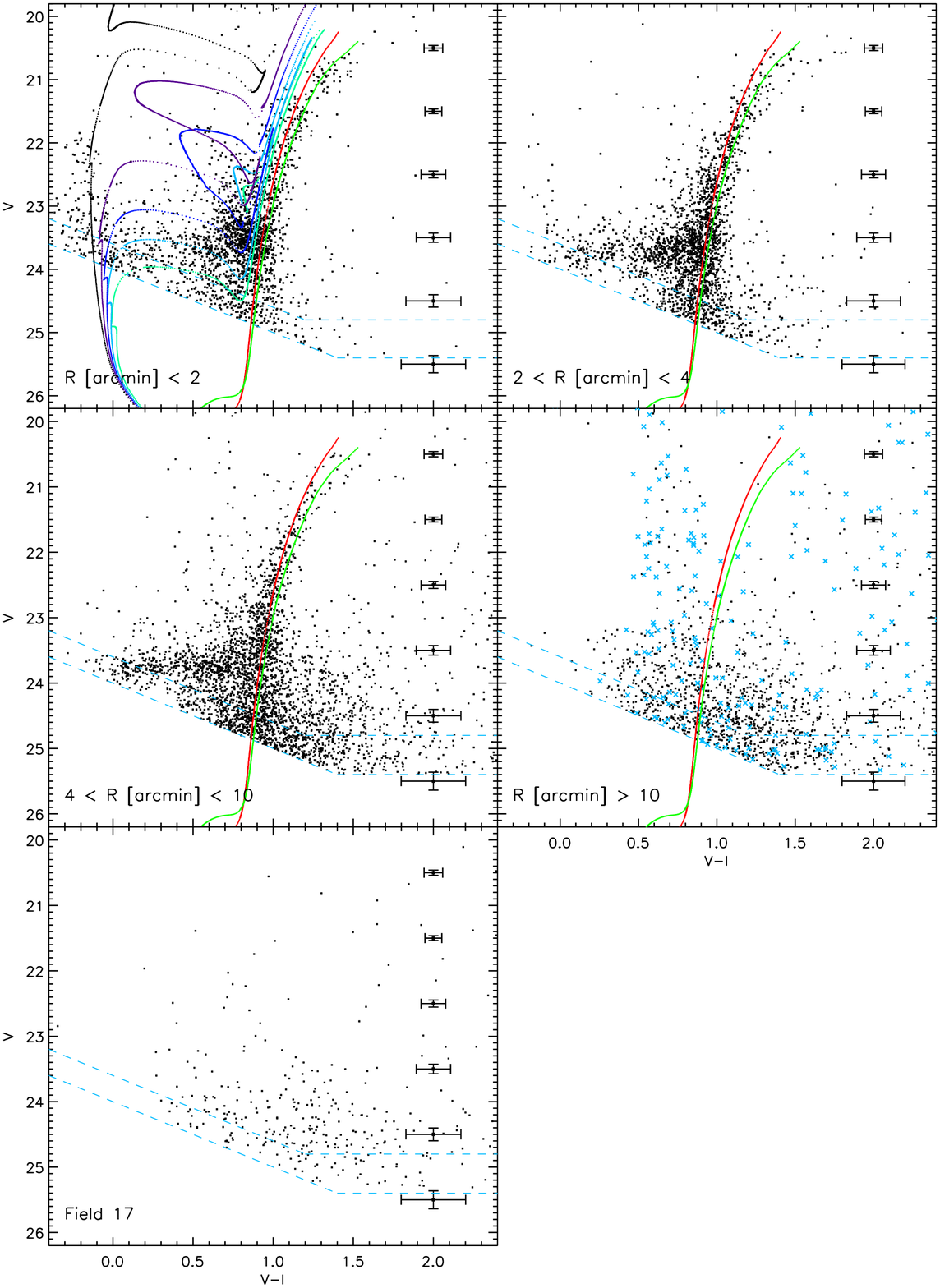}   
\caption{Colour-magnitude diagram for the objects classified as stellar in the Phoenix dwarf galaxy 
at different elliptical projected radii, R $<$ 2\arcmin, 2\arcmin $<$ R $<$ 4\arcmin, 
4\arcmin $<$ R $<$ 10\arcmin, R $>$ 10\arcmin\, and in the pointing displaced from the region of the mosaic (see labels). 
To allow for a meaningful comparison of the relative 
strenght of the various features in the CMD, the number of stars in the 
first 3 panels is  normalized, so that they contain the same number of stars above the S/N=5 
magnitude limit. In the last two panels, whose purpose is   
only to show where the foreground stars and unresolved galaxies are located on the CMD, we do not apply any 
normalization and show all the stars present. 
The dashed lines indicates the S/N= 5 and 10 limits. The squares  
with error-bars indicate the typical magnitude and color error per magnitude bins of 0.1 mag. BASTI 
isochrones \citep{pietrinferni2006} are 
overlaid on the CMD (red: 12 Gyr, alpha-enhanced, Z $= 3 \times 10^{-4}$; blue:   
8 Gyr, solar alpha, Z $= 6 \times 10^{-4}$; the other isochrones have Z $= 1 \times 10^{-3}$ and 
ages $=$ 0.1,0.3,0.5,0.7,1.0 Gyr from left to right). 
 The crosses on the panel R $>$ 10\arcmin show the location and amount 
of Milky Way foreground stars as predicted by the Besan\c{c}on model \citep{robin2003} 
onto the line-of-sight of Phoenix over an area 
corresponding to the elliptical annuli 10\arcmin $<$ R $<$ 13\arcmin, convolved with the observational errors in 
mag and color; the remaining objects show the location occupied by unresolved background galaxies within the observed CMDs. 
\label{fig:cmd_r}}
\end{figure*}

\begin{figure*}
\centering
\begin{tabular}{cc}
\includegraphics[width=0.4\linewidth]{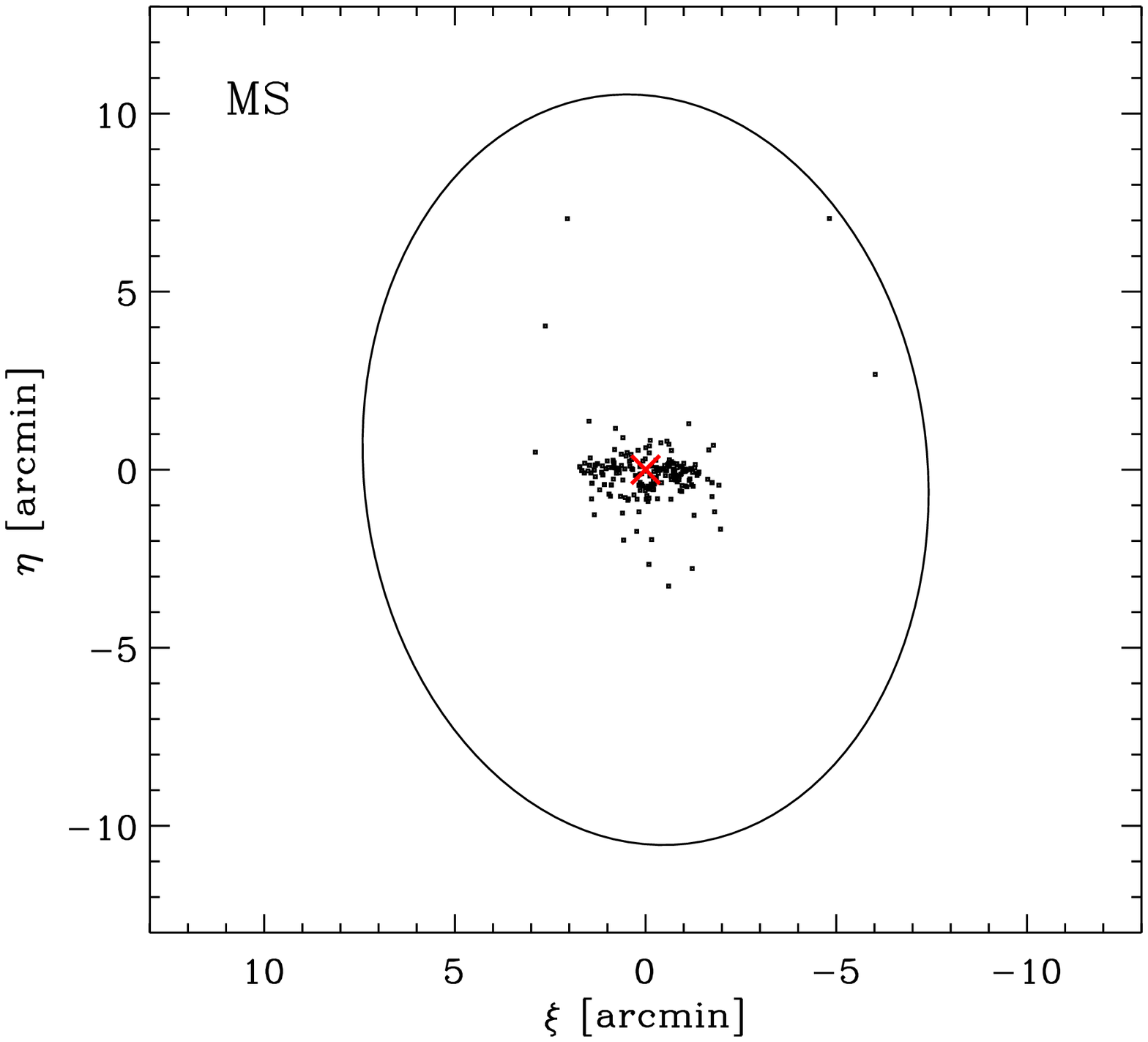} & 
\includegraphics[width=0.4\linewidth]{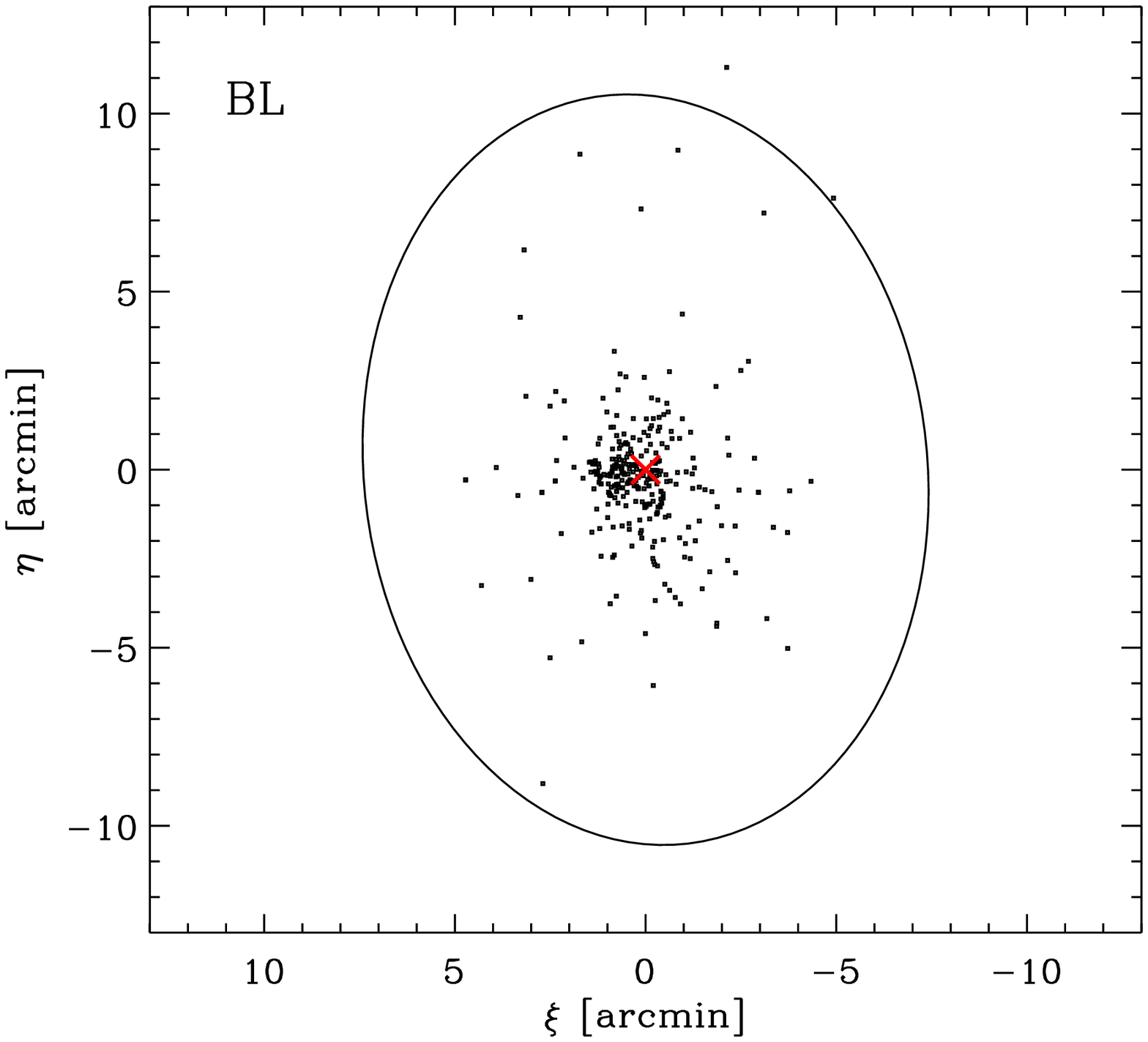}  \\ 
\includegraphics[width=0.4\linewidth]{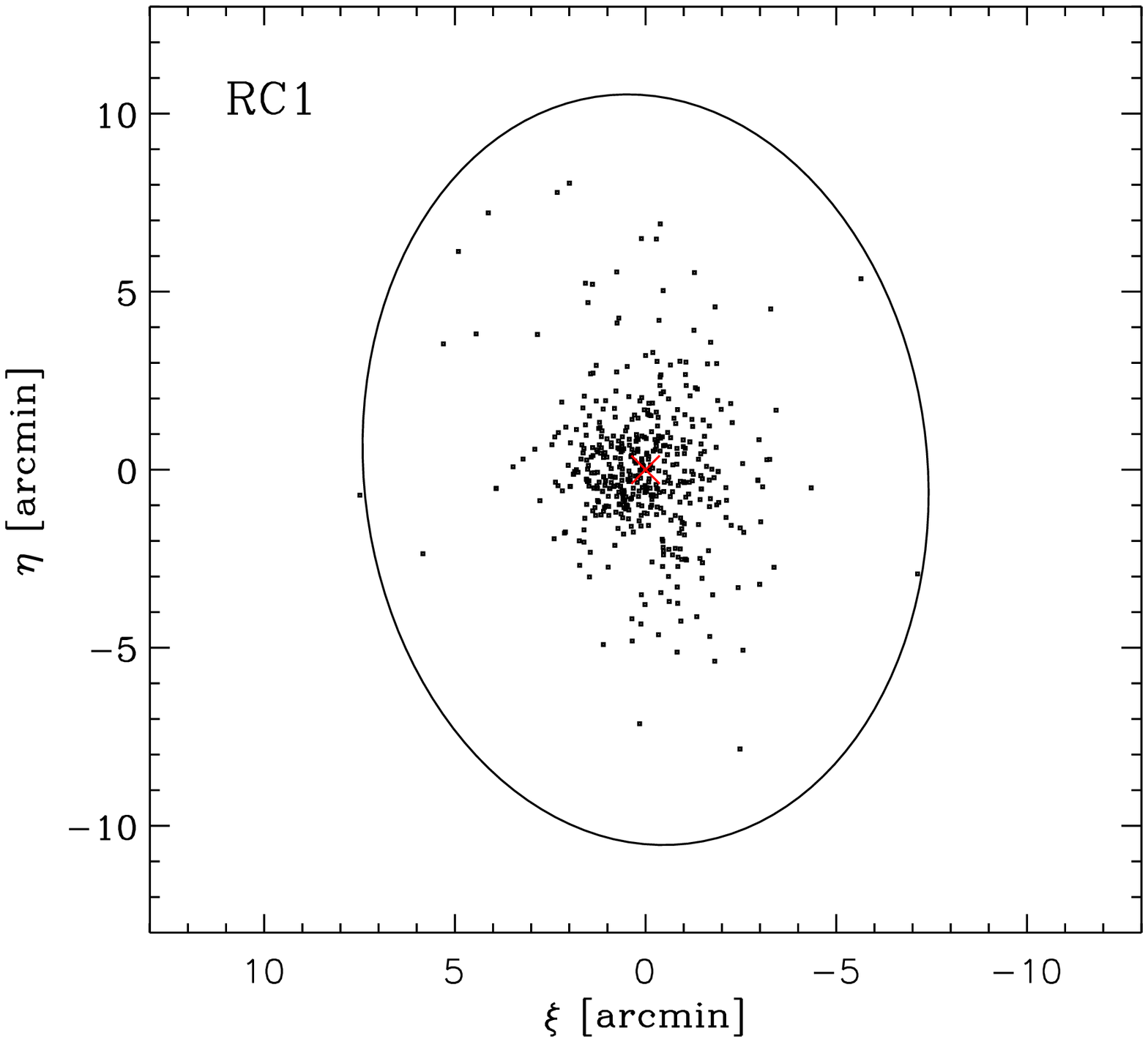} &  
\includegraphics[width=0.4\linewidth]{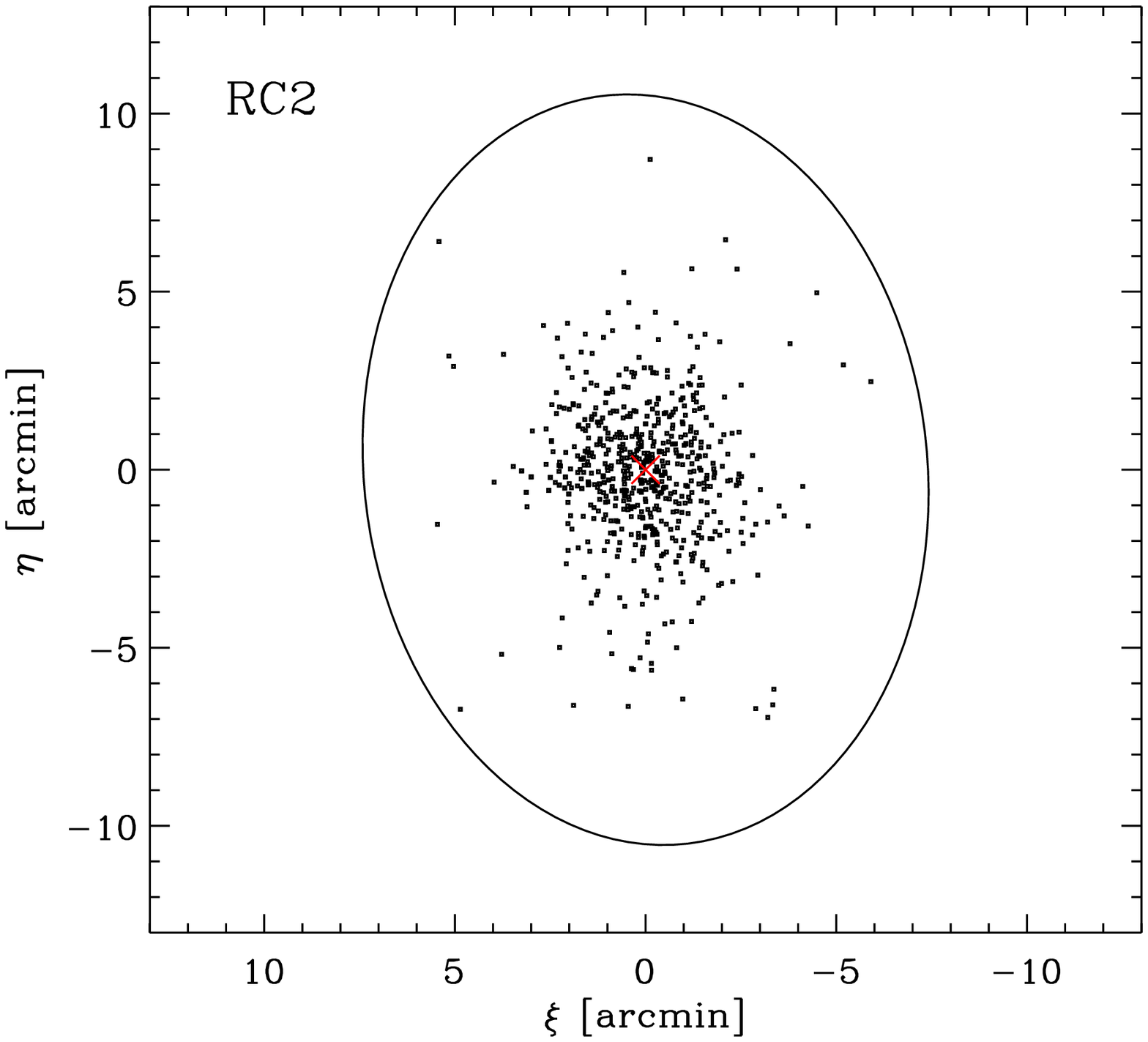} \\ 
\includegraphics[width=0.4\linewidth]{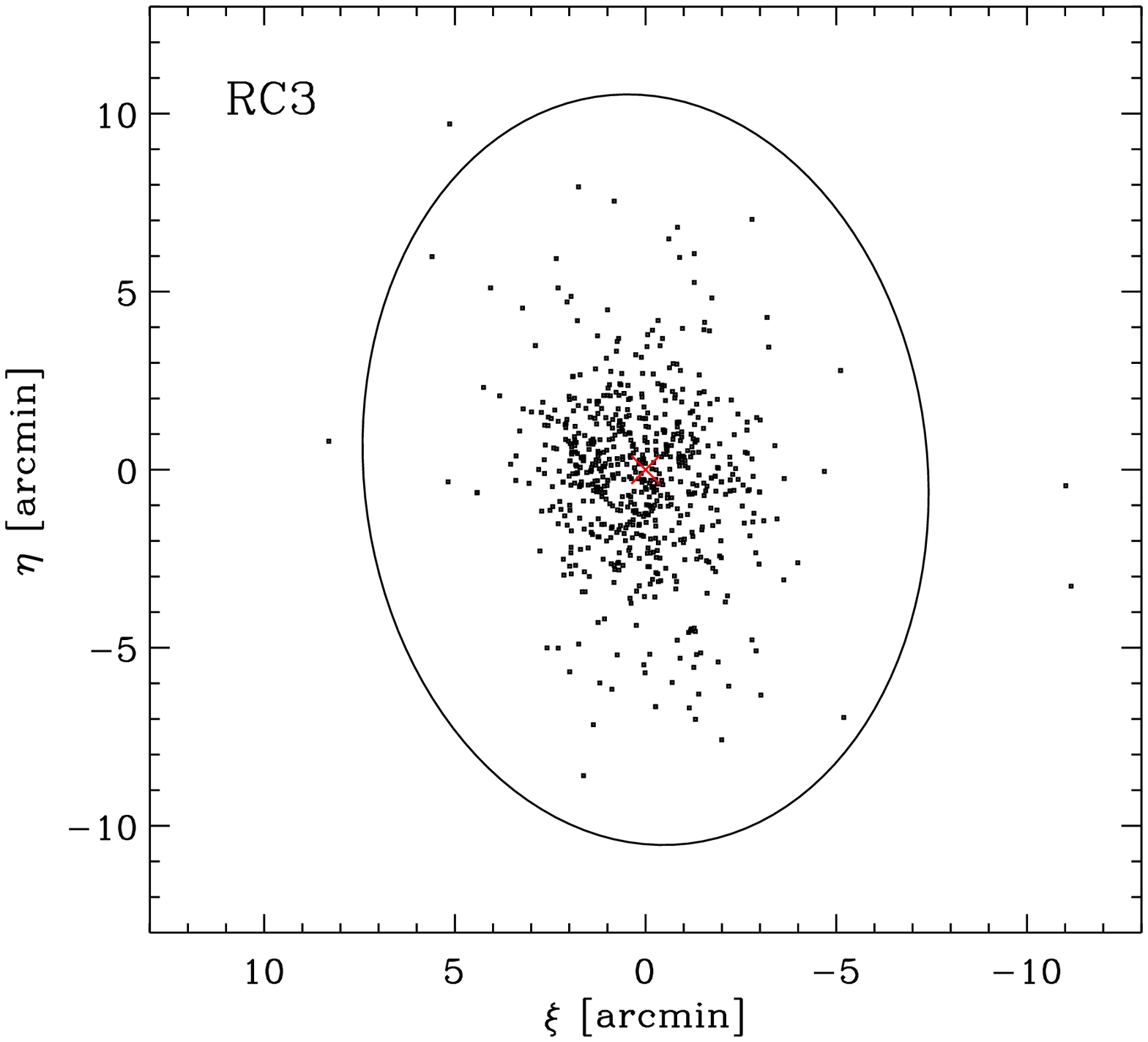} &  
\includegraphics[width=0.4\linewidth]{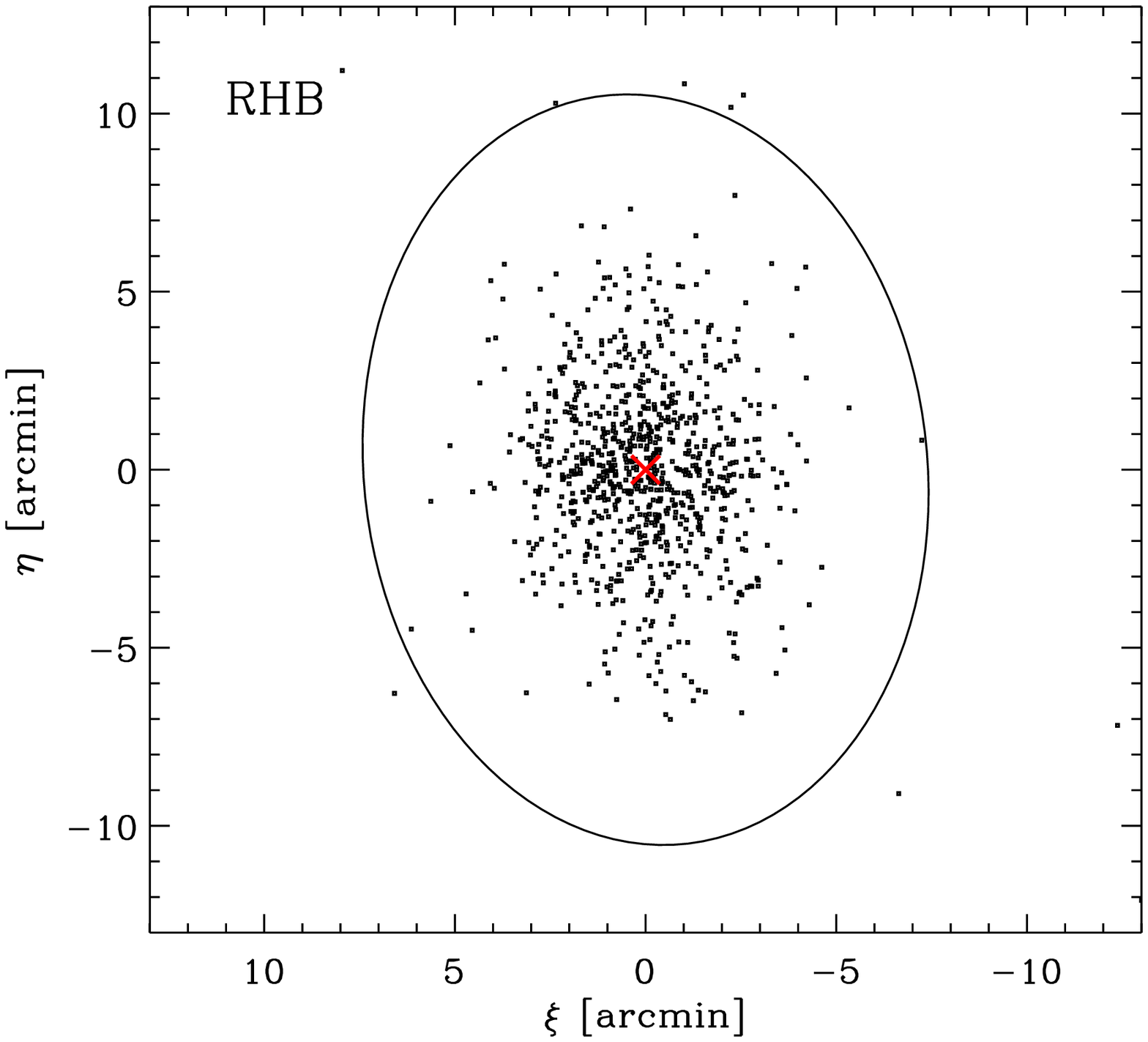} \\  
\end{tabular}
\caption{Spatial distribution of stars in different evolutionary stages and age ranges in the Phoenix dwarf galaxy, i.e. 
MS, BL, RC in bins of V magnitude 23.2-23.4 (RC1), 23.4-23.6 (RC2), 23.6-23.8 (RC3), and RHB stars as 
indicated by the labels and boxes in Fig.~\ref{fig:cmd}. 
The various populations have been extracted 
from the S/N$>$10 sample. A decontamination has been carried out, using the contaminant density derived 
as a weighted mean of the outer points of the surface density profile (except for the MS stars for which we used the  
contaminant density derived from the objects at $R>12.6$\arcmin). The ellipse shows the measurement  
of the nominal tidal radius derived in this work. A red cross is placed at the coordinates of Phoenix optical center to guide the eye. 
North is up and East is on the left. \label{fig:fov_pops_dec_wmean} }
\end{figure*}

A view of how the stellar population mix changes across the galaxy can already be gleaned by examining the 
CMD at different distances from the center. In Figure~\ref{fig:cmd_r} we plot CMDs in bins of projected elliptical 
radius: R $<$ 2\arcmin, 2\arcmin $<$ R $<$ 4\arcmin, 
4\arcmin $<$ R $<$ 10\arcmin. These CMDs are normalized to have the same number of stars, specifically  
the number of stars above the $V_{\rm 5}$, $I_{\rm 5}$ limit of the least populated spatial bin (4\arcmin $<$ R $<$ 10\arcmin). 
For comparison, the CMD at R $>$ 10\arcmin\,in the figure 
shows where the foreground stars and unresolved background galaxies should be located. The young MS and BL stars practically 
disappear at R $<$ 4\arcmin; the HB ($>$10 Gyr old) becomes much 
more enhanced going from the inner to the outer parts, and the RC becomes less extended in magnitude, 
indicating a decrease in the age range of stars in the stellar population mix; the latter is 
also confirmed by the tendency of the mean magnitude of RC stars in the various distance bins 
to become fainter at larger distances from the center, 
i.e. $I_{\rm RC} = 22.65, 22.75, 22.86$ and $V_{\rm RC} = 23.49, 23.60, 23.53$ 
for R $<$ 2\arcmin, 2\arcmin $<$ R $<$ 4\arcmin, 
4\arcmin $<$ R $<$ 10\arcmin, respectively.

Figure~\ref{fig:fov_pops_dec_wmean} shows the spatial distribution of 
stars sampling the different age ranges described in Sect.~\ref{sec:cmd}, i.e. MS, BL, 
RC in bins of V magnitude 23.2-23.4 (RC1), 23.4-23.6 (RC2), 23.6-23.8 (RC3), and RHB 
stars\footnote{We plot RHB stars rather than BHB because the latter are in smaller number and the selection 
box may suffer from contamination from the young MS stars.}. We carried out a decontamination from 
foreground stars and background galaxies in each of these plots by calculating the predicted number of 
contaminants on the area of the mosaic using the contaminant density derived from method a) applied to all these 
individual stellar populations, except for the MS stars to which we applied method b) because of their asymmetric 
distribution clearly different from the rest of the population (see the figure); in practise, the number of 
contaminants was randomly removed from the objects falling in a specific selection box, taking care of 
the fact that the density of contaminants is mostly expected to be uniform over such a small area. 

It is directly visible from Figure~\ref{fig:fov_pops_dec_wmean} 
that the younger the stellar population, the more centrally 
concentrated is its spatial distribution. This is particularly evident for the youngest stars, 
the MS and the BL, which also show a less extended distribution with respect to the rest. This would 
be consistent with the star formation region shrinking with time, with recent star formation 
having occurred in the innermost regions where the gas density would have still been high enough 
to sustain it (see also H09).

Such spatial variations of the stellar population mix had also been observed in previous studies using 
shallower and/or less spatially extended data \citep[e.g. MD99,][H09]{held1999}. The depth and spatial 
extent of our data allows us to perform a more accurate analysis than previous studies and 
quantify the variations in the spatial distribution of stars by deriving a surface 
number density profile for each population covering 
a different evolutionary stage (except for the flattened, asymmetric MS stars). The methodology and the 
functional forms fitted to the observed profiles are as described in Sect.~\ref{sec:surf}. The 
best-fitting parameters (see Table~\ref{tab:par_oldint}) show that the variations 
among the spatial distribution of the various stellar populations appear smooth, slowly and steadily 
going from more concentrated distributions from the youngest populations to more extended for the oldest. 
We checked that using the contaminant density derived from method b) to all stellar populations would 
bring negligible changes to the resulting spatial distributions, surface brightness profiles and resulting 
best-fit parameters (as in the case of the overall stellar population, the reduced $\chi^2$ values of the 
best-fitting profiles are larger when using method b) than method a) ). Also, the determination of the 
outer points over which we calculate the density of contaminants was adjusted by identifying where the 
surface number profile was becoming approximately flat; we checked that the choice of this region did not 
influence the results. 

As noted in previous works (e.g. MD99), the MS stars, those approximately younger than 0.7 Gyr, 
have a flattened distribution, almost perpendicular to the other stellar components. 
There appears to be two main clumps of young MS stars. Some clumping is also 
present in the distribution of the other young stars - the BL and the RC1 that represent the 
stars with ages mostly between 2-5 Gyr -, which both show a (rather spherical) 
concentration of stars displaced from the center, towards the South-East at 
rectangular coordinates (+0.5,-0.5); this is the direction of the 
coordinates we find for the position of the center in the inner 2\arcmin\,in Fig.~\ref{fig:par90}. 
Such displaced concentration 
disappears when moving to older ages, where the distribution becomes more regular. 
It can be noted that the populations on average older than 2 Gyr have a 
regular, spheroidal appearance and do not display a disk-like 
distribution as the young MS stars. Therefore the overall configuration of Phoenix 
does not resemble a disk-halo like larger spiral galaxies such as the MW: in the MW 
the disk contains stars of all ages (from young to ancient) while the halo is formed solely 
of ancient stars; this is different from the situation in Phoenix, where the disk-like feature is composed only of young stars, while the 
spheroidal part of the galaxy contains stars from intermediate to ancient ages.

We can speculate that the 
flattened, asymmetric distribution of the young MS stars still retains the imprint of where the 
recent star formation took place, and therefore of where the gas was recently located. This appears 
also to be still imprinted in the distribution of slightly older stars, such as the BL and the RC1, 
in the form of the clump displaced from the center. However, being slightly older than the MS, 
these BL and RC1 stars would have had more time to become sensitive to the overall potential 
of the galaxy, and for their spatial distribution to become smoother.

\section{Discussion} \label{sec:disc}

$\bullet$ {\it Spatial variations of the stellar population mix:} The spatial distribution of the stars changes with radius and becomes 
less and less centrally concentrated the older the stellar population. This is reflected in the values of the best-fitting parameters 
of the surface number profiles of the different stellar populations that steadily go from a more concentrated 
to less concentrated distribution with increasing ages. 

Similar variations in the stellar population mix appear to be a common 
characteristic of LG dwarf galaxies \citep[see e.g.][]{harbeck2001, tolstoy2004, battaglia2006, bernard2008, monelli2012}, 
with the star formation proceeding longer in the central regions. There are 
a number of mechanisms that could cause this: a natural evolution of the gas, 
for example progressively sinking in the center 
in absence of rotation \citep[e.g.][]{stinson2009, schroyen2011} or reaching higher densities in the center and therefore 
continuing star formation for longer. Another explanation could be
 ejection of the ISM because of supernovae explosions being more efficient in the 
outer parts; removal 
of the gas from ram pressure/tidal stripping \citep[e.g.][]{mayer2006}. 

\citet{stinson2009} show that age gradients of this type can arise  
in model isolated galaxies of low mass (with virial velocities $\la$20 \kms) without any need for external factors, 
because these systems do not form stable star forming discs, as the gas is mostly supported by 
pressure rather than angular momentum. Star formation rate is initially high in the center, because of the 
larger densities of the gas, and while the gas is consumed, the disc contracts until pressure support 
is re-established. Further shrinking may be due to a lower turbulence velocity due to the decline in SFR in the center, 
which mantains high gas densities in the centre but not in the outer regions, resulting in 
star formation that is systematically less radially extended as time goes on. 

It appears that whatever the mechanism is 
that creates evident 
variations in the stellar population mix, it can act on very different timescales from 
galaxy to galaxy. The Sculptor dSph, which predominantly consists of stars older than 
10 Gyr \citep[e.g.][]{deboer2012}, 
clearly displays this feature already in its ancient population \citep{tolstoy2004, deboer2012}. 
 In the Fornax dSph instead 
there is no detected spatial variation in the mix of 
the ancient population, while these changes are clear when comparing ancient, 
intermediate age and young stars \citep[e.g.][]{stetson1998, battaglia2006}. Phoenix in this respect is 
similar to Fornax. 
Such spatial variations between populations of different ages could be expected if they are due to environmental effects: 
more isolated objects and satellites with less internal/eccentric 
orbits around the host galaxy would be less affected and star formation could proceed longer in the outer parts; 
this would be consistent with what is presently known about the orbits of Sculptor and Fornax, bar the 
large error-bars in the proper motions values. At present there are no determinations of the proper motion 
of Phoenix; assuming a radial orbit for this object, its systemic velocity \citep[e.g.][]{irwin2002} 
referred to the Galactic Rest Frame (v$_{\rm sys, GSR} \sim -100$ \kms) would yield an r $\times$ v$^2$ about 1.5 times lower than for Leo~I, 
which is considered as a satellite of the Milky Way: in principle Phoenix could therefore still be a Milky Way satellite falling back. 
 
In order to disentangle environmental effects from internal ones, it would be instructive to quantify the 
differences in the distribution of stellar populations of various ages in a sample of LG dwarf galaxies. 
The isolated objects would give insights on the extent to which internal mechanisms can be responsible for 
the variations in the stellar population mix. It would be also instructive to compare systems with similar SFHs: 
for example Sculptor, Tucana and Cetus are all dSphs which produced most of their stars very early on, but 
while the former is a MW satellite, the other two are found far from the MW and M31. There are indications 
that also among the ancient stars in Tucana and Cetus, 
the older/metal-poor ones have a more extended spatial distribution than the younger/metal-rich 
\citep{harbeck2001, bernard2008, monelli2012}; it remains to be quantified how these differences in the distribution of the 
ancient stars compare between isolated and non-isolated dwarf galaxies. 

$\bullet$ {\it Spatial distribution of young stars:} The stars younger than 
0.7 Gyr display a clearly different spatial distribution than the rest 
of the population: not only 
they virtually disappear at distances larger than 4\arcmin\,from the center, but 
they are almost perpendicular to the bulk of the other stars, 
and the distribution is asymmetric and more flattened. This was already noted by VDK91 and MD99. 
By separating the sample in age bins, we also see that some asymmetries in the form of 
clumps are visible in the spatial distribution of BL stars and the younger RC stars, while 
stars older than about 5 Gyr display regular spatial distributions. 

We can speculate that the 
flattened, asymmetric distribution of the young MS stars still retains the imprint of where 
recent star formation took place, and therefore of where the gas was recently located. This appears 
also to be still imprinted in the distribution of the BL and young RC stars; however, since these 
stars are older than the young MS stars, 
they have had more time to diffuse to larger distance and to become more sensitive to the overall potential 
of the galaxy, resulting in a smoother distribution. Following this reasoning, it is likely that in a few 100s Myr 
also the young MS stars will diffuse to larger distances and acquire a more regular morphology, similar to the one 
of the main body of the galaxy. At that stage, Phoenix would probably look like a typical dSph, 
if it also were to lose its \hi gas or exhaust it in further star formation. 

In the models of Stinson et al. (2009) for low mass objects, although the majority of stars form systematically 
more inwards the more time passes, some amount of radial migration is also present, which 
goes in the direction of pushing stars to larger radii with respect to their birth site, 
qualitatively consistent with what we see here for the youngest stars.

We point out that a feature such as the one seen here - an inner flattened component containing young stars 
and tilted with respect to the main body of the galaxy - is also present in other LG dwarfs such as the 
Leo~A dIrr \citep[e.g. see Fig.~1 in][]{cole2007} and the closer by Fornax dSph, a MW satellite. With respect to 
the majority of MW dSphs, Fornax has had a much more extended SFH and is the only one to show stars possibly as young as 50 Myr 
\citep{coleman2008}. Also in this galaxy the young MS stars show a flattened 
distribution, tilted with respect to the main stellar population, while the slightly older BL stars 
have a less asymmetric and flattened distribution and appear to have diffused to larger distances. 
In this case, the tilting of the young stars is less enhanced, being approximately 40\degr \citep[e.g.][]{stetson1998}. 

It is unclear what would 
cause the tilting of the youngest stars with respect to the bulk of the stellar population. Such 
a behaviour has not been reported for more
massive, clearly rotating systems, such as for example the WLM dIrr \citep{leaman2012}, 
where the young stars would be aligned and concentric
with the distribution of older stars and the gas. 
This may indicate an influence of the total mass of the galaxy in determining its 
intrinsic angular momentum, and consequently the resulting distribution of its stellar populations: we 
speculate that, in less massive systems, the low amount of angular momentum would 
permit a disordered configuration of the gas, and therefore of 
the stars young enough not to have become sensitive to the overall potential of the galaxy. 
In apparent contradiction to that, in NGC~6822, a dIrr with a luminosity similar to WLM, the spatial distribution of 
the young stars and the old RGB stars are almost 
perpendicular to each other, with the young stars following the HI disk \citep[e.g.][]{demers2006}; 
in this case however the picture is complicated by the complex disk structure \citep{cannon2012} and the 
classification of NGC~6822 as a polar ring galaxy \citep[e.g.][]{demers2006}.

$\bullet$ {\it Disk-halo appearance:} 
Previous works (e.g. MD99) showed that the central parts of Phoenix display an inner component 
 that is tilted with respect to the main body of the galaxy and that 
morphologically resembles a disk - while the main body has a spheroidal 
distribution. The question was raised whether Phoenix may have a disk-halo structure similar to that of 
large spirals galaxies, such as the MW. 

We can comment on this issue simply making some consideration on the age of stars found 
in the ``disk''  and ``halo'' of Phoenix. 

In this work we showed that in Phoenix only young stars 
are responsible for the elongated central structure that morphologically resembles a disk, 
while such structure is absent for 
stars older than 2 Gyr, which display a rather regular  
spheroidal morphology. Therefore only stars younger than 1 Gyr are found in the ``disk'', while both intermediate age 
and ancient stars are found in the ``halo''. This is clearly different to what is observed for the MW, 
where the disk is known to contain stars of all ages while the halo consists of ancient stars; therefore the 
structure we see in Phoenix is not analogous to what found for the MW and M31.

 This result is also 
difficult to reconcile with the properties of stellar haloes formed in a $\Lambda$CDM context. 
Stellar haloes formed from disrupted satellite galaxies are a natural consequence of the hierarchical formation of galaxies 
in a $\Lambda$CDM framework and, for large spirals, the average age of stars in the stellar haloes is expected 
to be around 11 Gyr \citep{cooper2010}, with essentially no stars younger than 
5 Gyr, from mainly a few satellites accreted between redshift 1 and 3. Since in this context
 dwarf galaxies are expected to be among the first galactic systems to form, their stellar haloes should arguably have formed 
at earlier times than for large spirals, and from smaller progenitors, and should therefore contain only ancient stars.

\section{Summary and conclusions} \label{sec:concl}

We presented results from wide-area photometry in \mv and \mi\, band from VLT/FORS for Phoenix, 
one of the few transition type dwarf galaxies of the LG. 
The data consist of a mosaic of images covering 26\arcmin $\times$ 26\arcmin around the optical 
center of the galaxy and reaching below the horizontal branch of the system: this is 
the only data-set for this galaxy that combines relatively deep photometry with such a large spatial coverage. 
For comparison with previous studies, we can trace Phoenix stellar population out to projected radii of 1.5 kpc 
from its center, while the study of H09 reached out to 0.5kpc.

One output of the analysis is the re-determination of the overall structure of 
the system and especially of a determination of its extent, that was 
very uncertain in the literature. We derived a nominal tidal radius of 
$r_t =  10.56' \pm 0.15'$, about 60\% the value determined by MD99. The best-fitting profile 
to the overall stellar population of Phoenix is a Sersic profile of Sersic radius 
R$_S = 1.82' \pm 0.06'$ and $m=$0.83$\pm$0.03.

The distance of Phoenix has been derived independently from our dataset using the RGB tip method, 
and it is in agreement with previous distance determinations from H09 and M99. 
Examination of the CMD and LF 
revealed the presence of a bump above the red clump, consistent with
being a RGB bump. This is the first detection of this feature in Phoenix.

The depth of the photometry allows to study stars in different evolutionary phases, 
with ages ranging from 
about 0.1 Gyr to the oldest ages as they can be identified from features in the CMD. 
This, combined to the large area covered, enabled us to explore in a very direct way how the 
stellar population mix varies across the face of the galaxy. 

The spatial distribution of the stars changes with radius and becomes 
less and less centrally concentrated the older the stellar population. This is reflected in the values of the best-fitting parameters 
of the surface number profiles of the different stellar populations that we provide. 

As also found in previous studies, Phoenix displays in the inner regions a flattened structure, almost perpendicular to the main body of the 
galaxy. This feature was suggestive of a disk-halo structure and was raising the question whether also Phoenix could consist of a 
a disk-halo system such as the one found in the Milky Way.  
We showed that the flattened, tilted structure appears to 
be present mainly among stars younger than 1 Gyr, and absent for the stars $\ga$ 5 Gyr old, which on the other hand 
show a regular 
distribution also in the center of the galaxy. This argues against a disk-halo structure of the type 
found in large spirals such as the Milky Way. 

\section*{Acknowledgments}
The research leading to these results has received funding from the European Union
Seventh Framework Programme (FP7/2007-2013)
under grant agreement number PIEF-GA-2010-274151. We acknowledge the International Space Science Institute (ISSI) at Bern for their funding of the team 
``Defining the full life-cycle of dwarf galaxy evolution: the Local Universe as a template''. This work has made use of {\it BaSTI} web tools 
provided at www.oa-teramo.inaf.it/BASTI. We acknowlegde the use of the 
online catalogues for standard stars provided by P.Stetson at 
http://www3.cadc-ccda.hia-iha.nrc-cnrc.gc.ca/community/STETSON/standards/. GB thanks Michele Cignoni and Francesca Annibali for useful suggestions.

\bibliographystyle{mn2e} 
\bibliography{phoenix_biblio}

\begin{thebibliography}{}

\bibitem[\protect\citeauthoryear{{Akaike}}{{Akaike}}{1973}]{akaike1973}
{Akaike} H.,  1973 Second international symposium on information theory,
  {Information theory and an extension of the maximum likelihood principle}.
pp 267--281

\bibitem[\protect\citeauthoryear{{Appenzeller} \& {Rupprecht}}{{Appenzeller} \&
  {Rupprecht}}{1992}]{appenzeller1992}
{Appenzeller} I.,  {Rupprecht} G.,  1992, The Messenger, 67, 18

\bibitem[\protect\citeauthoryear{{Battaglia}, {Helmi}, {Tolstoy}, {Irwin},
  {Hill} \& {Jablonka}}{{Battaglia} et~al.}{2008}]{battaglia2008b}
{Battaglia} G.,  {Helmi} A.,  {Tolstoy} E.,  {Irwin} M.,  {Hill} V.,
  {Jablonka} P.,  2008, \apjl

\bibitem[\protect\citeauthoryear{{Battaglia}, {Tolstoy}, {Helmi}, {Irwin},
  {Letarte}, {Jablonka}, {Hill}, {Venn}, {Shetrone}, {Arimoto}, {Primas},
  {Kaufer}, {Francois}, {Szeifert}, {Abel} \& {Sadakane}}{{Battaglia}
  et~al.}{2006}]{battaglia2006}
{Battaglia} G.,  {Tolstoy} E.,  {Helmi} A.,  {Irwin} M.~J.,  {Letarte} B.,
  {Jablonka} P.,  {Hill} V.,  {Venn} K.~A.,  {Shetrone} M.~D.,  {Arimoto} N.,
  {Primas} F.,  {Kaufer} A.,  {Francois} P.,  {Szeifert} T.,  {Abel} T.,
  {Sadakane} K.,  2006, \aap, 459, 423

\bibitem[\protect\citeauthoryear{{Bellazzini}, {Beccari}, {Oosterloo},
  {Galleti}, {Sollima}, {Correnti}, {Testa}, {Mayer}, {Cignoni}, {Fraternali}
  \& {Gallozzi}}{{Bellazzini} et~al.}{2011}]{bellazzini2011}
{Bellazzini} M.,  {Beccari} G.,  {Oosterloo} T.~A.,  {Galleti} S.,  {Sollima}
  A.,  {Correnti} M.,  {Testa} V.,  {Mayer} L.,  {Cignoni} M.,  {Fraternali}
  F.,    {Gallozzi} S.,  2011, \aap, 527, A58

\bibitem[\protect\citeauthoryear{{Bellazzini}, {Gennari}, {Ferraro} \&
  {Sollima}}{{Bellazzini} et~al.}{2004}]{bellazzini2004}
{Bellazzini} M.,  {Gennari} N.,  {Ferraro} F.~R.,    {Sollima} A.,  2004,
  \mnras, 354, 708

\bibitem[\protect\citeauthoryear{{Bernard}, {Gallart}, {Monelli}, {Aparicio},
  {Cassisi}, {Skillman}, {Stetson}, {Cole}, {Drozdovsky}, {Hidalgo}, {Mateo} \&
  {Tolstoy}}{{Bernard} et~al.}{2008}]{bernard2008}
{Bernard} E.~J.,  {Gallart} C.,  {Monelli} M.,  {Aparicio} A.,  {Cassisi} S.,
  {Skillman} E.~D.,  {Stetson} P.~B.,  {Cole} A.~A.,  {Drozdovsky} I.,
  {Hidalgo} S.~L.,  {Mateo} M.,    {Tolstoy} E.,  2008, \apjl, 678, L21

\bibitem[\protect\citeauthoryear{{Caldwell}}{{Caldwell}}{1999}]{caldwell1999}
{Caldwell} N.,  1999, \aj, 118, 1230

\bibitem[\protect\citeauthoryear{{Cannon}, {O'Leary}, {Weisz}, {Skillman},
  {Dolphin}, {Bigiel}, {Cole}, {de Blok} \& {Walter}}{{Cannon}
  et~al.}{2012}]{cannon2012}
{Cannon} J.~M.,  {O'Leary} E.~M.,  {Weisz} D.~R.,  {Skillman} E.~D.,  {Dolphin}
  A.~E.,  {Bigiel} F.,  {Cole} A.~A.,  {de Blok} W.~J.~G.,    {Walter} F.,
  2012, ArXiv e-prints

\bibitem[\protect\citeauthoryear{{Caon}, {Capaccioli} \& {D'Onofrio}}{{Caon}
  et~al.}{1993}]{caon1993}
{Caon} N.,  {Capaccioli} M.,    {D'Onofrio} M.,  1993, \mnras, 265, 1013

\bibitem[\protect\citeauthoryear{{Cardelli}, {Clayton} \& {Mathis}}{{Cardelli}
  et~al.}{1989}]{cardelli1989}
{Cardelli} J.~A.,  {Clayton} G.~C.,    {Mathis} J.~S.,  1989, \apj, 345, 245

\bibitem[\protect\citeauthoryear{{Cole}, {Skillman}, {Tolstoy}, {Gallagher}
  III, {Aparicio}, {Dolphin}, {Gallart}, {Hidalgo}, {Saha}, {Stetson} \&
  {Weisz}}{{Cole} et~al.}{2007}]{cole2007}
{Cole} A.~A.,  {Skillman} E.~D.,  {Tolstoy} E.,  {Gallagher} III J.~S.,
  {Aparicio} A.,  {Dolphin} A.~E.,  {Gallart} C.,  {Hidalgo} S.~L.,  {Saha} A.,
   {Stetson} P.~B.,    {Weisz} D.~R.,  2007, \apjl, 659, L17

\bibitem[\protect\citeauthoryear{{Coleman}, {Da Costa} \&
  {Bland-Hawthorn}}{{Coleman} et~al.}{2005}]{coleman2005scl}
{Coleman} M.~G.,  {Da Costa} G.~S.,    {Bland-Hawthorn} J.,  2005, \aj, 130,
  1065

\bibitem[\protect\citeauthoryear{{Coleman}, {Da Costa}, {Bland-Hawthorn} \&
  {Freeman}}{{Coleman} et~al.}{2005}]{coleman2005}
{Coleman} M.~G.,  {Da Costa} G.~S.,  {Bland-Hawthorn} J.,    {Freeman} K.~C.,
  2005, \aj, 129, 1443

\bibitem[\protect\citeauthoryear{{Coleman} \& {de Jong}}{{Coleman} \& {de
  Jong}}{2008}]{coleman2008}
{Coleman} M.~G.,  {de Jong} J.~T.~A.,  2008, \apj, 685, 933

\bibitem[\protect\citeauthoryear{{Cooper}, {Cole}, {Frenk}, {White}, {Helly},
  {Benson}, {De Lucia}, {Helmi}, {Jenkins}, {Navarro}, {Springel} \&
  {Wang}}{{Cooper} et~al.}{2010}]{cooper2010}
{Cooper} A.~P.,  {Cole} S.,  {Frenk} C.~S.,  {White} S.~D.~M.,  {Helly} J.,
  {Benson} A.~J.,  {De Lucia} G.,  {Helmi} A.,  {Jenkins} A.,  {Navarro} J.~F.,
   {Springel} V.,    {Wang} J.,  2010, \mnras, 406, 744

\bibitem[\protect\citeauthoryear{{Da Costa} \& {Armandroff}}{{Da Costa} \&
  {Armandroff}}{1990}]{dacosta1990}
{Da Costa} G.~S.,  {Armandroff} T.~E.,  1990, \aj, 100, 162

\bibitem[\protect\citeauthoryear{{de Boer}, {Tolstoy}, {Hill}, {Saha}, {Olsen},
  {Starkenburg}, {Lemasle}, {Irwin} \& {Battaglia}}{{de Boer}
  et~al.}{2012}]{deboer2012}
{de Boer} T.~J.~L.,  {Tolstoy} E.,  {Hill} V.,  {Saha} A.,  {Olsen} K.,
  {Starkenburg} E.,  {Lemasle} B.,  {Irwin} M.~J.,    {Battaglia} G.,  2012,
  \aap, 539, A103

\bibitem[\protect\citeauthoryear{{Demers}, {Battinelli} \& {Kunkel}}{{Demers}
  et~al.}{2006}]{demers2006}
{Demers} S.,  {Battinelli} P.,    {Kunkel} W.~E.,  2006, \apjl, 636, L85

\bibitem[\protect\citeauthoryear{{Freudling}, {Romaniello}, {Patat},
  {M{\o}ller}, {Jehin} \& {O'Brien}}{{Freudling} et~al.}{2007}]{freudling2007}
{Freudling} W.,  {Romaniello} M.,  {Patat} F.,  {M{\o}ller} P.,  {Jehin} E.,
  {O'Brien} K.,  2007, in {C.~Sterken} ed., The Future of Photometric,
  Spectrophotometric and Polarimetric Standardization Vol.~364 of Astronomical
  Society of the Pacific Conference Series, {Photometry with FORS at the ESO
  VLT}.
p.~113

\bibitem[\protect\citeauthoryear{{Gallart}, {Aparicio}, {Freedman}, {Madore},
  {Mart{\'{\i}}nez-Delgado} \& {Stetson}}{{Gallart} et~al.}{2004}]{gallart2004}
{Gallart} C.,  {Aparicio} A.,  {Freedman} W.~L.,  {Madore} B.~F.,
  {Mart{\'{\i}}nez-Delgado} D.,    {Stetson} P.~B.,  2004, \aj, 127, 1486

\bibitem[\protect\citeauthoryear{{Gallart}, {Mart{\'{\i}}nez-Delgado},
  {G{\'o}mez-Flechoso} \& {Mateo}}{{Gallart} et~al.}{2001}]{gallart2001}
{Gallart} C.,  {Mart{\'{\i}}nez-Delgado} D.,  {G{\'o}mez-Flechoso} M.~A.,
  {Mateo} M.,  2001, \aj, 121, 2572

\bibitem[\protect\citeauthoryear{{Gallart}, {Zoccali} \& {Aparicio}}{{Gallart}
  et~al.}{2005}]{gallart2005}
{Gallart} C.,  {Zoccali} M.,    {Aparicio} A.,  2005, \araa, 43, 387

\bibitem[\protect\citeauthoryear{{Girardi}, {Bressan}, {Bertelli} \&
  {Chiosi}}{{Girardi} et~al.}{2000}]{girardi2000}
{Girardi} L.,  {Bressan} A.,  {Bertelli} G.,    {Chiosi} C.,  2000, \aaps, 141,
  371

\bibitem[\protect\citeauthoryear{{Girardi} \& {Salaris}}{{Girardi} \&
  {Salaris}}{2001}]{girardi2001}
{Girardi} L.,  {Salaris} M.,  2001, \mnras, 323, 109

\bibitem[\protect\citeauthoryear{{Graham} \& {Guzm{\'a}n}}{{Graham} \&
  {Guzm{\'a}n}}{2003}]{graham2003}
{Graham} A.~W.,  {Guzm{\'a}n} R.,  2003, \aj, 125, 2936

\bibitem[\protect\citeauthoryear{{Harbeck}, {Grebel}, {Holtzman},
  {Guhathakurta}, {Brandner}, {Geisler}, {Sarajedini}, {Dolphin},
  {Hurley-Keller} \& {Mateo}}{{Harbeck} et~al.}{2001}]{harbeck2001}
{Harbeck} D.,  {Grebel} E.~K.,  {Holtzman} J.,  {Guhathakurta} P.,  {Brandner}
  W.,  {Geisler} D.,  {Sarajedini} A.,  {Dolphin} A.,  {Hurley-Keller} D.,
  {Mateo} M.,  2001, \aj, 122, 3092

\bibitem[\protect\citeauthoryear{{Held}, {Saviane} \& {Momany}}{{Held}
  et~al.}{1999}]{held1999}
{Held} E.~V.,  {Saviane} I.,    {Momany} Y.,  1999, \aap, 345, 747

\bibitem[\protect\citeauthoryear{{Hidalgo}, {Aparicio},
  {Mart{\'{\i}}nez-Delgado} \& {Gallart}}{{Hidalgo} et~al.}{2009}]{hidalgo2009}
{Hidalgo} S.~L.,  {Aparicio} A.,  {Mart{\'{\i}}nez-Delgado} D.,    {Gallart}
  C.,  2009, \apj, 705, 704

\bibitem[\protect\citeauthoryear{{Holtzman}, {Smith} \& {Grillmair}}{{Holtzman}
  et~al.}{2000}]{holtzman2000}
{Holtzman} J.~A.,  {Smith} G.~H.,    {Grillmair} C.,  2000, \aj, 120, 3060

\bibitem[\protect\citeauthoryear{{Irwin} \& {Hatzidimitriou}}{{Irwin} \&
  {Hatzidimitriou}}{1995}]{IH1995}
{Irwin} M.,  {Hatzidimitriou} D.,  1995, \mnras, 277, 1354

\bibitem[\protect\citeauthoryear{{Irwin} \& {Tolstoy}}{{Irwin} \&
  {Tolstoy}}{2002}]{irwin2002}
{Irwin} M.,  {Tolstoy} E.,  2002, \mnras, 336, 643

\bibitem[\protect\citeauthoryear{{Jerjen} \& {Rejkuba}}{{Jerjen} \&
  {Rejkuba}}{2001}]{jerjen2001}
{Jerjen} H.,  {Rejkuba} M.,  2001, \aap, 371, 487

\bibitem[\protect\citeauthoryear{{Karachentsev}, {Karachentseva}, {Huchtmeier}
  \& {Makarov}}{{Karachentsev} et~al.}{2004}]{karachentsev2004}
{Karachentsev} I.~D.,  {Karachentseva} V.~E.,  {Huchtmeier} W.~K.,    {Makarov}
  D.~I.,  2004, \aj, 127, 2031

\bibitem[\protect\citeauthoryear{{Kazantzidis}, {{\L}okas}, {Callegari},
  {Mayer} \& {Moustakas}}{{Kazantzidis} et~al.}{2011}]{kazantzidis2011}
{Kazantzidis} S.,  {{\L}okas} E.~L.,  {Callegari} S.,  {Mayer} L.,
  {Moustakas} L.~A.,  2011, \apj, 726, 98

\bibitem[\protect\citeauthoryear{{King}}{{King}}{1962}]{king1962}
{King} I.,  1962, \aj, 67, 471

\bibitem[\protect\citeauthoryear{{Leaman}, {Venn}, {Brooks}, {Battaglia},
  {Cole}, {Ibata}, {Irwin}, {McConnachie}, {Mendel} \& {Tolstoy}}{{Leaman}
  et~al.}{2012}]{leaman2012}
{Leaman} R.,  {Venn} K.~A.,  {Brooks} A.~M.,  {Battaglia} G.,  {Cole} A.~A.,
  {Ibata} R.~A.,  {Irwin} M.~J.,  {McConnachie} A.~W.,  {Mendel} J.~T.,
  {Tolstoy} E.,  2012, \apj, 750, 33

\bibitem[\protect\citeauthoryear{{Lee}, {Freedman} \& {Madore}}{{Lee}
  et~al.}{1993}]{lee1993}
{Lee} M.~G.,  {Freedman} W.~L.,    {Madore} B.~F.,  1993, \apj, 417, 553

\bibitem[\protect\citeauthoryear{{Majewski}, {Frinchaboy}, {Kunkel}, {Link},
  {Mu{\~n}oz}, {Ostheimer}, {Palma}, {Patterson} \& {Geisler}}{{Majewski}
  et~al.}{2005}]{majewski2005}
{Majewski} S.~R.,  {Frinchaboy} P.~M.,  {Kunkel} W.~E.,  {Link} R.,
  {Mu{\~n}oz} R.~R.,  {Ostheimer} J.~C.,  {Palma} C.,  {Patterson} R.~J.,
  {Geisler} D.,  2005, \aj, 130, 2677

\bibitem[\protect\citeauthoryear{{Marigo}, {Girardi}, {Bressan}, {Groenewegen},
  {Silva} \& {Granato}}{{Marigo} et~al.}{2008}]{marigo2008}
{Marigo} P.,  {Girardi} L.,  {Bressan} A.,  {Groenewegen} M.~A.~T.,  {Silva}
  L.,    {Granato} G.~L.,  2008, \aap, 482, 883

\bibitem[\protect\citeauthoryear{{Mart{\'{\i}}nez-Delgado},
  {Alonso-Garc{\'{\i}}a}, {Aparicio} \&
  {G{\'o}mez-Flechoso}}{{Mart{\'{\i}}nez-Delgado} et~al.}{2001}]{MD2001}
{Mart{\'{\i}}nez-Delgado} D.,  {Alonso-Garc{\'{\i}}a} J.,  {Aparicio} A.,
  {G{\'o}mez-Flechoso} M.~A.,  2001, \apjl, 549, L63

\bibitem[\protect\citeauthoryear{{Mart{\'{\i}}nez-Delgado}, {Gallart} \&
  {Aparicio}}{{Mart{\'{\i}}nez-Delgado} et~al.}{1999}]{martinez1999}
{Mart{\'{\i}}nez-Delgado} D.,  {Gallart} C.,    {Aparicio} A.,  1999, \aj, 118,
  862

\bibitem[\protect\citeauthoryear{{Mateo}}{{Mateo}}{1998}]{mateo1998}
{Mateo} M.~L.,  1998, \araa, 36, 435

\bibitem[\protect\citeauthoryear{{Mayer}, {Mastropietro}, {Wadsley}, {Stadel}
  \& {Moore}}{{Mayer} et~al.}{2006}]{mayer2006}
{Mayer} L.,  {Mastropietro} C.,  {Wadsley} J.,  {Stadel} J.,    {Moore} B.,
  2006, \mnras, 369, 1021

\bibitem[\protect\citeauthoryear{{Menzies}, {Feast}, {Whitelock}, {Olivier},
  {Matsunaga} \& {da Costa}}{{Menzies} et~al.}{2008}]{menzies2008}
{Menzies} J.,  {Feast} M.,  {Whitelock} P.,  {Olivier} E.,  {Matsunaga} N.,
  {da Costa} G.,  2008, \mnras, 385, 1045

\bibitem[\protect\citeauthoryear{{Monelli}, {Bernard}, {Gallart}, {Fiorentino},
  {Drozdovsky}, {Aparicio}, {Bono}, {Cassisi}, {Skillman} \&
  {Stetson}}{{Monelli} et~al.}{2012}]{monelli2012}
{Monelli} M.,  {Bernard} E.~J.,  {Gallart} C.,  {Fiorentino} G.,  {Drozdovsky}
  I.,  {Aparicio} A.,  {Bono} G.,  {Cassisi} S.,  {Skillman} E.~D.,
  {Stetson} P.~B.,  2012, ArXiv e-prints

\bibitem[\protect\citeauthoryear{{Monelli}, {Cassisi}, {Bernard}, {Hidalgo},
  {Aparicio}, {Gallart} \& {Skillman}}{{Monelli} et~al.}{2010}]{monelli2010}
{Monelli} M.,  {Cassisi} S.,  {Bernard} E.~J.,  {Hidalgo} S.~L.,  {Aparicio}
  A.,  {Gallart} C.,    {Skillman} E.~D.,  2010, \apj, 718, 707

\bibitem[\protect\citeauthoryear{{Pietrinferni}, {Cassisi}, {Salaris} \&
  {Castelli}}{{Pietrinferni} et~al.}{2004}]{pietrinferni2004}
{Pietrinferni} A.,  {Cassisi} S.,  {Salaris} M.,    {Castelli} F.,  2004, \apj,
  612, 168

\bibitem[\protect\citeauthoryear{{Pietrinferni}, {Cassisi}, {Salaris} \&
  {Castelli}}{{Pietrinferni} et~al.}{2006}]{pietrinferni2006}
{Pietrinferni} A.,  {Cassisi} S.,  {Salaris} M.,    {Castelli} F.,  2006, \apj,
  642, 797

\bibitem[\protect\citeauthoryear{{Plummer}}{{Plummer}}{1911}]{plummer1911}
{Plummer} H.~C.,  1911, \mnras, 71, 460

\bibitem[\protect\citeauthoryear{{Robin}, {Reyl{\'e}}, {Derri{\`e}re} \&
  {Picaud}}{{Robin} et~al.}{2003}]{robin2003}
{Robin} A.~C.,  {Reyl{\'e}} C.,  {Derri{\`e}re} S.,    {Picaud} S.,  2003,
  \aap, 409, 523

\bibitem[\protect\citeauthoryear{{Salaris}, {Chieffi} \& {Straniero}}{{Salaris}
  et~al.}{1993}]{salaris1993}
{Salaris} M.,  {Chieffi} A.,    {Straniero} O.,  1993, \apj, 414, 580

\bibitem[\protect\citeauthoryear{{Schlegel}, {Finkbeiner} \&
  {Davis}}{{Schlegel} et~al.}{1998}]{schlegel1998}
{Schlegel} D.~J.,  {Finkbeiner} D.~P.,    {Davis} M.,  1998, \apj, 500, 525

\bibitem[\protect\citeauthoryear{{Schroyen}, {de Rijcke}, {Valcke},
  {Cloet-Osselaer} \& {Dejonghe}}{{Schroyen} et~al.}{2011}]{schroyen2011}
{Schroyen} J.,  {de Rijcke} S.,  {Valcke} S.,  {Cloet-Osselaer} A.,
  {Dejonghe} H.,  2011, \mnras, 416, 601

\bibitem[\protect\citeauthoryear{{Sersic}}{{Sersic}}{1968}]{sersic1968}
{Sersic} J.~L.,  1968, {Atlas de galaxias australes}.
Cordoba, Argentina: Observatorio Astronomico, 1968

\bibitem[\protect\citeauthoryear{{Stanek} \& {Garnavich}}{{Stanek} \&
  {Garnavich}}{1998}]{stanek1998}
{Stanek} K.~Z.,  {Garnavich} P.~M.,  1998, \apjl, 503, L131

\bibitem[\protect\citeauthoryear{{Stetson}}{{Stetson}}{1987}]{stetson1987}
{Stetson} P.~B.,  1987, \pasp, 99, 191

\bibitem[\protect\citeauthoryear{{Stetson}}{{Stetson}}{2000}]{stetson2000}
{Stetson} P.~B.,  2000, \pasp, 112, 925

\bibitem[\protect\citeauthoryear{{Stetson}, {Hesser} \&
  {Smecker-Hane}}{{Stetson} et~al.}{1998}]{stetson1998}
{Stetson} P.~B.,  {Hesser} J.~E.,    {Smecker-Hane} T.~A.,  1998, \pasp, 110,
  533

\bibitem[\protect\citeauthoryear{{Stinson}, {Dalcanton}, {Quinn}, {Gogarten},
  {Kaufmann} \& {Wadsley}}{{Stinson} et~al.}{2009}]{stinson2009}
{Stinson} G.~S.,  {Dalcanton} J.~J.,  {Quinn} T.,  {Gogarten} S.~M.,
  {Kaufmann} T.,    {Wadsley} J.,  2009, \mnras, 395, 1455

\bibitem[\protect\citeauthoryear{{Tolstoy}, {Hill} \& {Tosi}}{{Tolstoy}
  et~al.}{2009}]{tolstoy2009}
{Tolstoy} E.,  {Hill} V.,    {Tosi} M.,  2009, \araa, 47, 371

\bibitem[\protect\citeauthoryear{{Tolstoy}, {Irwin}, {Helmi}, {Battaglia},
  {Jablonka}, {Hill}, {Venn}, {Shetrone}, {Letarte}, {Cole}, {Primas},
  {Francois}, {Arimoto}, {Sadakane}, {Kaufer}, {Szeifert} \& {Abel}}{{Tolstoy}
  et~al.}{2004}]{tolstoy2004}
{Tolstoy} E.,  {Irwin} M.~J.,  {Helmi} A.,  {Battaglia} G.,  {Jablonka} P.,
  {Hill} V.,  {Venn} K.~A.,  {Shetrone} M.~D.,  {Letarte} B.,  {Cole} A.~A.,
  {Primas} F.,  {Francois} P.,  {Arimoto} N.,  {Sadakane} K.,  {Kaufer} A.,
  {Szeifert} T.,    {Abel} T.,  2004, 617, L119

\bibitem[\protect\citeauthoryear{{Trujillo}, {Erwin}, {Asensio Ramos} \&
  {Graham}}{{Trujillo} et~al.}{2004}]{trujillo2004}
{Trujillo} I.,  {Erwin} P.,  {Asensio Ramos} A.,    {Graham} A.~W.,  2004, \aj,
  127, 1917

\bibitem[\protect\citeauthoryear{{van de Rydt}, {Demers} \& {Kunkel}}{{van de
  Rydt} et~al.}{1991}]{vdk1991}
{van de Rydt} F.,  {Demers} S.,    {Kunkel} W.~E.,  1991, \aj, 102, 130

\bibitem[\protect\citeauthoryear{{Wilkinson}, {Kleyna}, {Evans}, {Gilmore},
  {Irwin} \& {Grebel}}{{Wilkinson} et~al.}{2004}]{wilkinson2004}
{Wilkinson} M.~I.,  {Kleyna} J.~T.,  {Evans} N.~W.,  {Gilmore} G.~F.,  {Irwin}
  M.~J.,    {Grebel} E.~K.,  2004, \apjl, 611, L21

\bibitem[\protect\citeauthoryear{{Young}, {Skillman}, {Weisz} \&
  {Dolphin}}{{Young} et~al.}{2007}]{young2007}
{Young} L.~M.,  {Skillman} E.~D.,  {Weisz} D.~R.,    {Dolphin} A.~E.,  2007,
  \apj, 659, 331

\end{thebibliography}


\appendix


\section{RGB and AGB bump}

\begin{figure*}
\includegraphics[width=\linewidth]{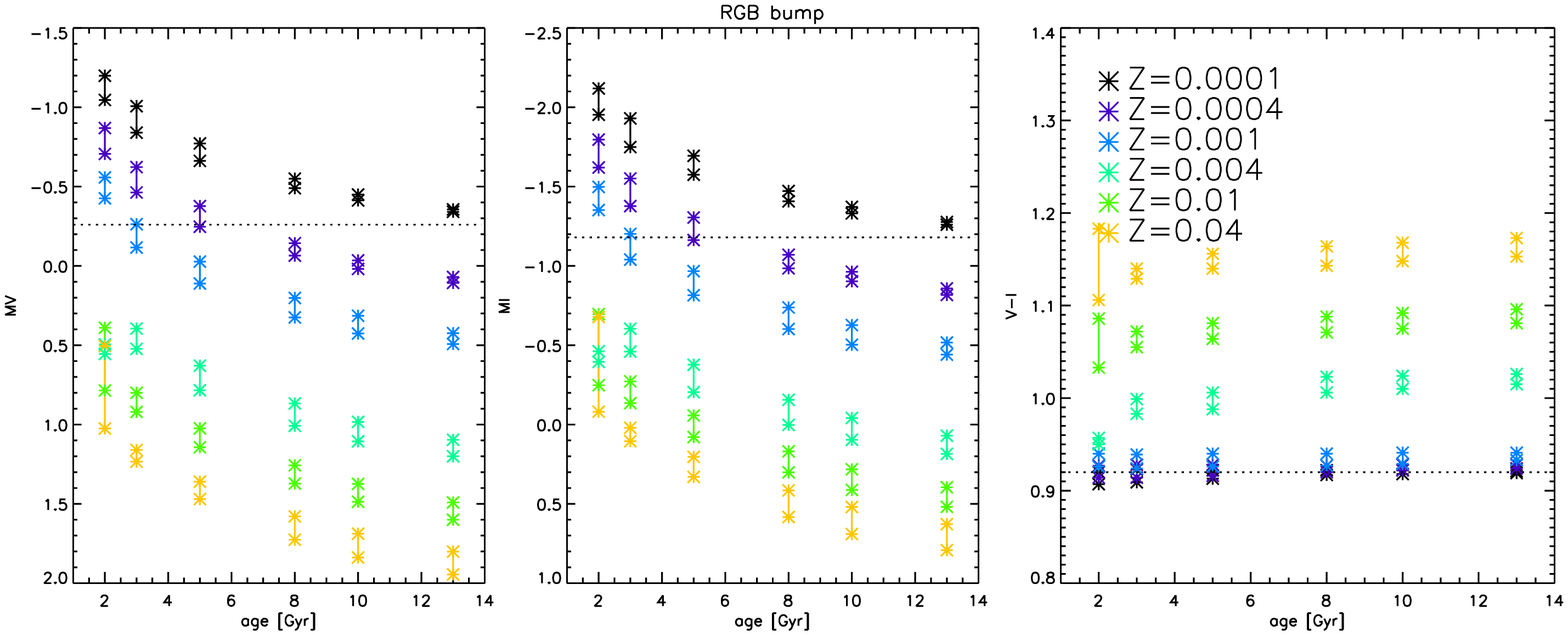}
\includegraphics[width=\linewidth]{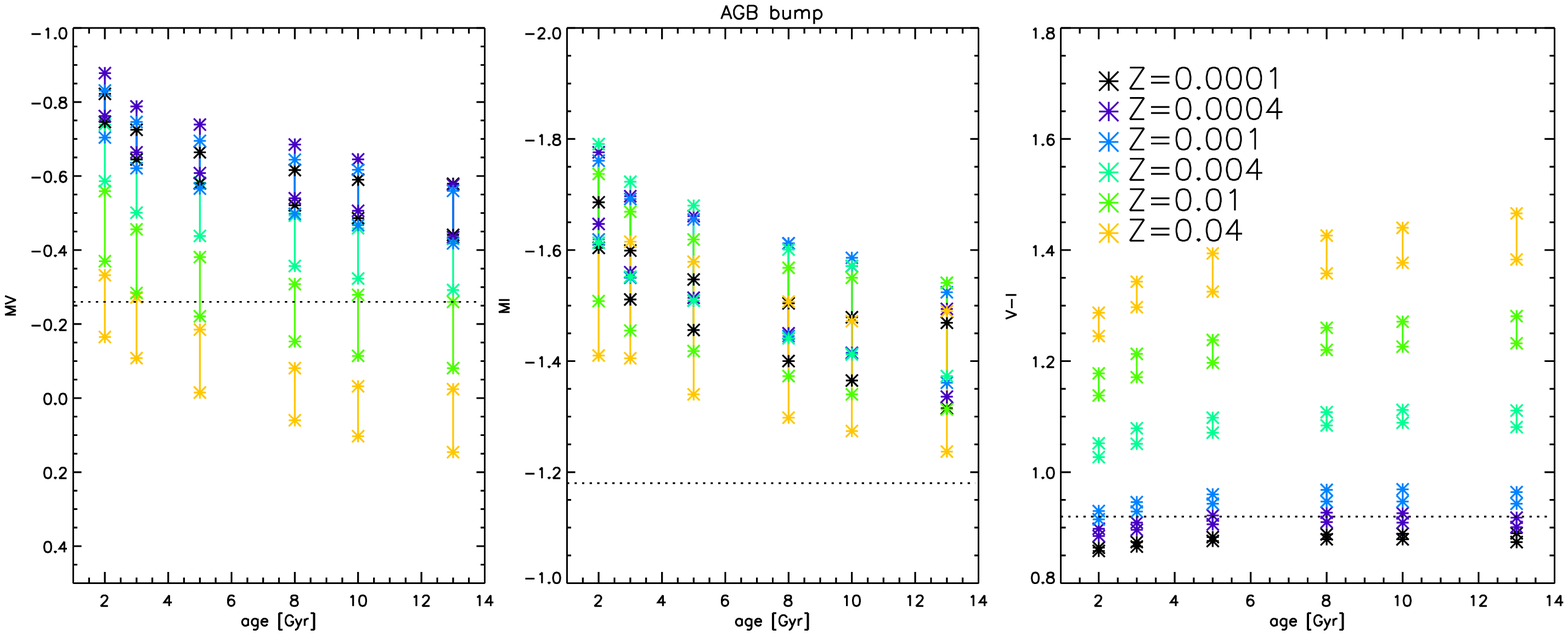}
\caption{For this figure we used solar-scaled Padua isochrones (Marigo et al. 2008) spanning an age range from 2 to 13 Gyr, Z=0.0001-0.02. 
Top: The symbols connected by vertical lines show the luminosity interval 
	of RGB stars crossing the discontinuity in chemical 
	profile left by the first dredge-up event. This interval 
	corresponds to the bump in the luminosity function along the 
	RGB. Bottom: as before but this interval 
	corresponds to the clump of early-AGB stars in 
	colour-magnitude diagrams, therefore the AGB bump. 
        From left to right we show how the V mag, I mag and \vi vary as a function 
        of age in Gyr for isochrones of different metallicities. 
} \label{fig:bump}
\end{figure*}


\bsp

\label{lastpage}

\end{document}